%% file: main.tex
\documentclass[acmsmall,nonacm]{acmart}
\usepackage{multirow}
\usepackage[export]{adjustbox}
\usepackage{booktabs}
\usepackage{siunitx}
\usepackage{enumitem}
\usepackage{color}
\usepackage{subcaption}
\usepackage{hyperref}
\usepackage{tabularx}
\usepackage{graphicx}
\usepackage{caption}

\AtBeginDocument{%
  \providecommand\BibTeX{{%
    \normalfont B\kern-0.5em{\scshape i\kern-0.25em b}\kern-0.8em\TeX}}}

\newcommand{\xhdr}[1]{\vspace{1.7mm}\noindent{{\bf #1.}}}

\newcommand{\celltext}[1]{\noindent{{\footnotesize{\textit{#1}}}}}

\newcommand{\celltextpl}[1]{\noindent{{\footnotesize{#1}}}}

\setcopyright{acmcopyright}
\copyrightyear{2023}
\acmYear{2023}
\acmDOI{10.1145/1122445.1122456}

\usepackage{enumitem}
\usepackage{booktabs}
\usepackage{multirow}

\newcommand\RQ[1]{\textbf{#1}}
\newif\ifshowcomments
\showcommentsfalse

\ifshowcomments
\fi

\begin{document}

\title{Rhythm of Work: Mixed-methods Characterization of Information Workers Scheduling Preferences and Practices}


\author{Lu Sun}
\email{l5sun@ucsd.edu}
\affiliation{%
  \institution{University of California, San Diego}
  \city{La Jolla}
  \state{California}
  \country{USA}
}

\author{Lillio Mok}
\email{lillio@cs.toronto.edu}
\affiliation{%
  \institution{University of Toronto}
  \city{Toronto}
  \state{Ontario}
  \country{Canada}
}

\author{Shilad Sen}
\email{shilad.sen@microsoft.com}
\affiliation{%
  \institution{Microsoft}
  \city{Redmond}
  \state{Washington}
  \country{USA}
}
\affiliation{%
  \institution{Macalester College}
  \city{St. Paul}
  \state{Minnesota}
  \country{USA}
}

\author{Bahareh Sarrafzadeh}
\email{basarraf@microsoft.com}
\affiliation{%
  \institution{Microsoft}
  \city{Redmond}
  \state{Washington}
  \country{USA}
}


\begin{abstract}
As processes around hybrid work, spatially distant collaborations, and work-life boundaries grow increasingly complex, managing workers' schedules for synchronous meetings has become a critical aspect of building successful global teams.
However, gaps remain in our understanding of workers' scheduling preferences and practices, which we aim to fill in this large-scale, mixed-methods study of individuals calendars in a multinational organization.
Using interviews with eight participants, survey data from 165 respondents, and telemetry data from millions of meetings scheduled by 211 thousand workers, we characterize scheduling preferences, practices, and their relationship with each other and organizational factors.
We find that temporal preferences can be broadly classified as either cyclical, such as suitability of certain days, or relational, such as disperse meetings, at various time scales.
Furthermore, our results suggest that these preferences are disconnected from actual practice~--~albeit with several notable exceptions~--~and that individual differences are associated with factors like meeting load, time-zones, importance of meetings to job function, and job titles.
We discuss key themes for our findings, along with the implications for calendar and scheduling systems and socio-technical systems more broadly.
\end{abstract}

\ccsdesc[500]{Human-centered computing~Empirical studies in HCI}

\keywords{work day, calendar preference, information workers} 


\maketitle

\input{1-Introduction.tex}
\input{2-Background.tex}
\input{3-FormativeStudy.tex}

\input{4-Method.tex}

\input{5.1-Results-Preferences.tex}

\input{5.2-Results-Practices.tex}

\input{6-Discussion.tex}

\input{7-Limitation.tex}




\bibliographystyle{ACM-Reference-Format}
\bibliography{ref}
\input{Appendix.tex}

\end{document}

%% file: 1-Introduction.tex
\section{Introduction}

Organizations are increasingly concerned with fostering effective global collaborations as they grow larger, spread across the world, and become disperse~\cite{neeley2015global}.
A central aspect of facilitating this now-distributed work, but previously centralized, is the need for collaborators to meet synchronously.
Media synchronicity theory, for instance, argues that synchronous communication is needed for convergence upon a shared understanding of information~\cite{dennis2008media,dennis1999rethinking}. 
Empirically, meetings enable rapid feedback and reduce delays in coordination~\cite{olson2000distance,tang2011your,cummings2009crossing,espinosa2006effect,espinosa2011time}, without which teams must wait for information and expertise from collaborators~\cite{agerfalk2005framework,morrison2020challenges,cummings2011geography}.
Furthermore, workers who can interact synchronously may feel more motivated~\cite{dourish1992awareness, morrison2020challenges}, more easily develop trust in one another~\cite{sarker2011role,choi2019mechanism,dube2009surviving}, and be exposed to diverse viewpoints~\cite{batarseh2017collaboration,mok2023challenging}.

In spite of their benefits, however, the task of \textit{scheduling} meetings remains complex and requires accommodating multiple kinds of temporal constraints.
For one, people anecdotally display an inherent ``rhythm'' in how they organize work around daily events, such as lunch breaks~\cite{reddy2002finger,jackson2011collaborative}.
Additionally, there are inherent trade-offs between time spent meeting with colleagues and time spent carrying out individual work duties, leading to calendaring strategies that protect time for personal work~\cite{tietze2003times,jalote2004timeboxing}. 
The prioritization of time away from meetings extends further to workers' offline needs, such as familial obligations~\cite{williams2023managing,devault1991feeding}. 
Because meetings involve multiple collaborators, workers also need to negotiate mutually-acceptable timeslots and find ``common ground'' between attendees~\cite{cramton2001mutual, dourish1992awareness}, an endeavor that is even more difficult when colleagues are geographically and temporally distant~\cite{olson2000distance,mok2023challenging}.
Thus, to optimize schedules, information workers need to be aware of work rhythms, capture daily preferences, and identify the discrepancies between constraints and preferences to concretely shape their workday~\cite{berry2007balancing, oh2005calendar}.

Calendaring behaviors at work have been further complicated by the COVID-19 pandemic, which radically altered organizations to adopt temporally flexible work rhythms~\cite{breideband2022home} and physically flexible remote collaboration modalities~\cite{ahrendt2020living,teevan2022microsoft}.
These shifts have correspondingly supported flexibility in how individuals' lives at home and work are separated, such that workers no longer have to adhere strictly to a ``traditional'' 9-to-5 workday with potentially beneficial outcomes~\cite{shifrin2022flexible}.
And yet, longer workdays~\cite{MSWTI, awada2021working} and more variance in working hours~\cite{MSTriplePeak,morshed2022stress} may also leave workers with less time for completing tasks or taking breaks~\cite{yang2022effects}.

How can these existing and novel challenges of scheduling be tackled for an era of post-pandemic work?
Despite a rich body of work on the design of intelligent systems for capturing temporal needs~\cite{dent1992personal,mitchell1994experience,sen1997satisfying,chajewska2000making,tullio2002augmenting,oh2005calendar,gervasio2005active,brzozowski2006grouptime,krzywicki2010adaptive,kim2018learning}, many of these advances are yet untested \textit{in situ}.
Indeed, we find that they are mostly guided either by self-reported scheduling practices or by hypothetical, simulated scheduling agents.
More fundamentally, the literature has also yet to develop a generic catalogue of the times during which workers prefer to meet.
Thus, the foundational question of how workers \textit{empirically} schedule in practice and whether their behaviors organically align with temporal preferences remains unclear. 
Answering this question is necessary not only to inform the design of calendaring systems, but also to guide organizational policies on worker schedules and time management more broadly.

In this study, we aim to fill the gaps in our understanding of scheduling practices by studying information workers' calendars in a large, multinational organization.
We pose two research questions to focus our work: 
\begin{itemize}
    \item {\RQ{RQ1-Preferences:} What are the main categories of preference around meeting arrangement? }
    \item {\RQ{RQ2-Practices:} How are these preferences aligned with the existing calendar practices? }
\end{itemize}

To address these questions, we structure our mixed-methods investigation as follows. 
We begin with a set of interviews within a contextual inquiry \textbf{formative study}, from which we derive a framework for broadly categorizing workers' temporal preferences.
We then conduct our \textbf{main study} informed by insights from the formative study, comprising of three parts. 
First, we utilize a set of $n=165$ survey responses to uncover how these preferences vary across different members of the organization.
Second, we use a \texttt{Large Telemetry Dataset} which contains telemetric trace of $6.9$ million meetings scheduled at the company to characterize the degree to which temporal preferences are aligned with workers' actual calendars.
Lastly, we delve deeper into the alignment between workers' stated preferences and their actual scheduling practices use the \texttt{Survey-Linked Telemetry Dataset} by linking telemetry data with survey responses from 101 consenting survey participants.
Throughout, organizational factors that contribute to differences in preferences, practices, and their alignment are interrogated and delineated. 

\xhdr{Summary of results}
We surface two key classes of temporal preferences around scheduling that represent \textit{cyclical} preferences, such as time of day, and \textit{relational} preferences, such as how clustered or dispersed meetings fall within a day. 
In terms of cyclical time periods, workers generally prefer to schedule meetings away from the edges of time periods; however, relational preferences are less universal and more personal.
Factor analysis revealed that workers' commute requirements, meeting load, and the nature of tasks influenced workers' preference for their meeting time. In addition, a higher meeting load may lead workers to prioritize clustering meetings on specific days when they arrange meetings. 
Comparing workers' preferred meeting time versus their actual meeting time in the survey linked telemetry dataset, we further find that scheduling practices rarely reflect general cyclical preferences~--~although workers in some circumstances appear able to schedule during preferred times, such as when they are meeting organizers.  
Through our telemetry analysis, broadly, organizational factors that interact with these patterns include the meeting load workers have, the timezone in which they reside, their roles, and the perceived importance of meetings to their job functions. 
Meeting load in practice limits users' choices of when to take breaks, whereas high meeting load users have fewer variations on when to take breaks. 

This study thus contributes a novel, generic framework for understanding the rhythm of synchronous, collaborative meetings in workers' calendars.
It also teases apart the subtleties in workers' stated preferences across different time scales and different organizational factors, which can be used to guide more effective calendaring behaviors in distributed teams.
By demonstrating that, observationally, scheduling behaviors are misaligned with preferences, our results emphasizes the importance of empirically grounding system design for and policy making around employees' schedules.

%% file: 2-Background.tex
\section{Background}
Successful time management is a key aspect of human well-being and is crucial for our biological~\cite{healy2008breaks}, cognitive~\cite{schmidt2007time}, and social~\cite{romero2017influence, bodker2013calendars} functions.
Organizationally, effective scheduling is even more salient for increasingly complex and globalized businesses, whose employees must navigate myriad factors to arrange time in their calendars for meeting and collaborating~\cite{cao2021large,mok2023challenging,hongladarom2002web}.
Without finding time for meetings, workers would be disafforded an immediate way of exchanging information~\cite{mell2021bridging,mok2023challenging} that would allow collaborators to converge on common ground~\cite{dennis1999rethinking} and reduce delays in coordination~\cite{olson2000distance, espinosa2011time}.
Indeed, the intricacies of workplace scheduling have been further complicated by shifts towards remote work induced by COVID-19~\cite{teevan2022microsoft, yang2022effects}, blurring the boundary between work and personal life~\cite{allen2021boundary,hakkila2020practical}.

Our study clarifies the complex preferences and behaviors underpinning how information workers schedule, and is situated against two bodies of related research delineated in this section.
First, we survey the literature on \textit{work rhythms} and the temporal patterns according to which people undertake personal and labor tasks.
Second, we examine research that aim to design calendaring systems by capturing anecdotal and theoretical accounts of workers' temporal preferences.

\subsection{The rhythm of work}
Information workers exhibit rhythm in their daily life, including when they usually start the day, attend meetings and take lunch breaks~\cite{reddy2002finger,jackson2011collaborative,breideband2022home}. 
These rhythms play a vital role in workdays by helping to maintain working productivity and balance work and personal life~\cite{tang2001connexus,begole2002work,begole2003rhythm,mark2014bored}. 
Here, we examine key aspects of these rhythms that can influence work patterns and organizational outcomes. 
Firstly, a rich body of CSCW work has explored rhythms of work related to the \textbf{day of week} in different contexts, including doctors, developers, scientific researchers, and information workers~\cite{begole2002work,reddy2002finger,begole2003rhythm,jackson2011collaborative,mark2014bored,meyer2019today,breideband2022home}. 
Previous research found that some developers organize their day of the week based on the activities they usually conduct, such as bug-fixing day, meeting day, or planning day~\cite{meyer2019today}. 
Developers reported that they usually scheduled ``no meeting days'' where they reserve one weekday, such as Monday, to focus on coding and development. 
Researchers proposed a hypothesis on the ``Blue Monday effect'', where people are in bad mood on Mondays, which affects the ability to focus on work~\cite{stone1985prospective}. 
Faced with this hypothesis, a previous in-situ tracking study with information workers found that people report Monday being the most bored, but at the same time also the most focused~\cite{mark2014bored}. 

Research has also found that \textbf{circadian rhythms} and people's inner clock have strong influences on people's daily activities, including sleep, work productivity, and mood~\cite{cajochen2003role,murray2010circadian,steinhardt2014reconciling,jun2019circadian,janbocke2020finding,raiha2021evening,breideband2022home}. 
Previous research used chronotypes to describe individuals' circadian preferences, such as ``early birds'' or ``night owls''~\cite{jun2019circadian,schmidt2007time}. 
Morning chronotypes often awaken, prefer to start working, and feel tired earlier in the day. Others are ``night owls'' who experience the patterns later in the day and tend to awaken later, stay up working, and go to bed later. 
Growing evidence showed that circadian rhythms impact individual cognition and productivity as well as group collaborative work~\cite{jun2019circadian,wieth2011time,folkard2003shift,schmidt2007time}. 
Researchers even developed calendar applications that incorporate chronotype information to help people manage their day~\cite{janbocke2020finding}. 

The modality of work also influences work rhythms. 
For instance, \textbf{remote or hybrid workers} with families play multiple social roles in life \cite{grimes2008life}. 
They need to not only manage schedules with teams but also coordinate with other family members on responsibilities, such as picking up children from daycare. 
Previous CSCW work investigated the usage of calendars with family and found that individuals used calendars to build shared understanding and reduce tensions~\cite{elliot2005awareness,hakkila2020practical,neustaedter2009calendar,bodker2013calendars}. 
Remote modalities are also determined by whether co-workers are \textbf{geographically disperse}, which makes it more difficult for collaborators to capture and maintain rhythmic patterns~\cite{olson2000distance}. 
In turn, lack of awareness of rhythms in a distributed team increases the cost of coordinating meetings and collaborations~\cite{begole2003rhythm,jackson2011collaborative,morrison2020challenges}.
When teams are also \textbf{temporally distanced} across different timezones, meetings become significantly harder to schedule but more important for keeping collaborations diverse and well-connected~\cite{mok2023challenging}.

These considerations have become even more salient during the \textbf{COVID-19 pandemic}, during which companies have shifted towards infrastructure and business models that enable employees to work remotely and from home~\cite{defilippis2020collaborating,yang2022effects}. 
Studies have shown that workdays have become longer~\cite{MSWTI} and blurred with personal life~\cite{allen2021boundary,hakkila2020practical}. 
A previous study on workers found that the 9-to-5 workday is fading in an age of remote and hybrid work and some people have grown accustomed to the ``triple peak day'', where there is a third peak raised after 9-to-5 workday hours~\cite{MSTriplePeak}. 
It raised the question that whether workers perceive this triple peak as more flexible or as an encroachment on personal hours that are normally used for family time or social life. 
Simultaneously, there are also signs that teams may have become less productive~\cite{gibbs2021work}, less focused~\cite{cao2021large}, and less connected~\cite{yang2022effects}. 

The rhythms are therefore viewed as important factors in helping balance \textbf{productivity and well-being}. 
Workers often block time on their calendars to reduce distractions and prevent interruptions \cite{mark2018effects}. 
An interruption at the wrong time, e.g. when users are in a state of flow, can result in lower task productivity and increased frustration~\cite{mihaly1997finding,mark2008cost,mcternan2013depression}. However, long continuous working hours  without breaks can induce more mental stress, so people always take breaks due to tiredness and boredom \cite{epstein2016taking,strongman2000taking}. Previous research showed that work breaks, such as stepping away from one’s work area to get food or drink, can contribute positively to workers' mental and physical well-being and improve their creativity and productivity~\cite{healy2008breaks,epstein2016taking,cambo2017breaksense,mark2018effects,kaur2020optimizing}.

\subsection{Capturing scheduling preferences}\label{section:bg:calendar}

It is therefore unsurprising, given the complex temporal dynamics governing people's workday, that scheduling is a complex personal behavior. 
Whether the user is initiating a new meeting or responding to a meeting request she chooses an action with multiple objectives in mind~\cite{berry2007balancing,oh2005calendar}. 
For instance, when trying to schedule a new meeting, the user may want to minimize change to her existing meetings, and she takes a scheduling action that best compromises the overall objectives, such as: minimizing travel time, maximizing a contiguous meeting time, choosing a preferred meeting location, etc. 
As a result, a calendar scheduling system that tries to predict the best scheduling actions needs to know user's preferences and priorities \cite{oh2005calendar}.

Learning and incorporating personal preference in calendar scheduling systems received previous attention~\cite{dent1992personal,mitchell1994experience,sen1997satisfying,chajewska2000making,tullio2002augmenting,oh2005calendar,gervasio2005active,brzozowski2006grouptime,krzywicki2010adaptive,kim2018learning}. 
\citeauthor{mitchell1994experience} proposed Calendar Apprentice (CAP) which is a decision tree based calendar manager that can learn decision rules for predicting the values of schedule attributes such as day, time and location~\cite{mitchell1994experience}. 
PLIANT used an active learning approach where the learning system can learn user preferences from the feedback that naturally occurs during interactive scheduling \cite{gervasio2005active}. 

To understand personal preferences, scheduling systems must deal with challenges in  preference elicitation.
The user may not be fully aware of their preferences or be able to express them~\cite{berry2007balancing}. 
Even in cases where the user does know about their preferences and is able to express them, she may not be willing to input them into the system, especially if the elicitation process is burdensome. 
The first approach to the elicitation of general scheduling preferences aims to derive rankings of schedule options by showing user-specific examples as part of an elicitation phase. 
GroupTime~\cite{brzozowski2006grouptime} is a scheduling system that takes such an example-driven approach. 
\citeauthor{viappiani2006evaluating} show that presenting users with carefully chosen examples can help stimulate their preferences, a technique known as example critiquing~\cite{viappiani2006evaluating}. 
An alternative approach to elicitation requires the user to enter the parameters of the preference model directly. 
The scheduling system of \citeauthor{haynes1997automated} takes such a model-driven approach. The advantage of this approach is that the elicitation period can be shorter because the information elicited has greater entropy \cite{haynes1997automated}. 
The significant disadvantage is that users must comprehend the model, rather than examples that are possibly easier to comprehend.

Some scheduling systems took the user-centered approach to understand and elicit user preferences. 
In PTIME \cite{berry2007balancing}, they first conduct a structured interview  with subjects to capture their criteria on scheduling and used the findings of the user study as well as features suggested in prior work in the literature (e.g., \cite{kozierok1993learning,mitchell1994experience,brzozowski2006grouptime}), to identify seven criteria consistent across different users, such as duration of meetings and scheduling windows for the requested meetings.
Later, the system provided an interactive visual interface that present several statements around scheduling criteria such as ``avoid rescheduling'' or ``meeting duration'' and ask participants to rank the importance of each criterion. 
However, this system provided general criteria for the meeting within the days and there are still a large number of preference categories, such as ``meeting gaps'', that are not captured. 

\xhdr{Relation to this work}
Together, this rich body of research paints a nuanced picture of how an individual's workday can be influenced by various temporal preferences and necessities, and how these preferences can be distilled into guidance for designing scheduling aids.
However, several gaps remain in our understanding of these calendaring dynamics, which we address at present.

Firstly, while the literature identifies several situation-specific preferences (e.g. Mondays~\cite{stone1985prospective,mark2014bored} and no-meeting days~\cite{meyer2019today}), we still lack a generalized taxonomy of temporal needs capturing different time scales and temporal relationships like breaks.
Furthermore, the relationship between self-reported, stated calendaring preferences and what workers do in practice remains unclear.
Although employees in large organizations may not easily meet their scheduling preferences, it is unknown whether their actual calendars are populated with sub-optimal timeslots~--~nor, indeed, what their actual calendars look like in general.
Finally, the major workplace changes induced by COVID-19 cast much uncertainty over established observations of scheduling behaviors. 
On the one hand, increasingly flexible workdays may allow calendaring at preferable hours; on the other, the blending of work and non-work days may impinge upon times reserved for offline obligations.

Our work thus contributes a mixed-methods contextual inquiry and survey to disentangle scheduling preferences, complimented by a large-scale characterization of workers' \textit{in situ} calendaring behaviors.
This endeavors to fill the empirical gaps in our understanding of how calendars can be better organized, and is the first to do so across the entirety of a multinational organization in the post-pandemic paradigm of work. 

%% file: 3-FormativeStudy.tex
\section{Formative Study}\label{section:formative}
In order to gain initial insights into users' preferences regarding optimal meeting arrangements on their calendars, as well as to explore the influencing factors behind their scheduling decisions, we conducted a series of contextual inquiries with employees from a major technology company. These inquiries involved discussions about their job roles, calendaring practices, the types of meetings they commonly schedule or attend, and the time management challenges they encounter.

\subsection{Participants}
Specifically, we ran eight exploratory interviews, averaging 45 minutes, focusing on eliciting an ideal arrangement of events on participants’ calendars and factors that impact their scheduling decisions. 
We recruited a group of 8 employees with diverse job roles within the technology company, ranging from product managers to chiefs of staff. Among the participants, 6 out of 8 were female, while 2 out of 8 were male. Their tenure at the company varied from 3 to 8 years.
Table \ref{tab:InterviewRoles} summarizes our participants roles at the company.

\begin{table}[t]
    \centering
    \begin{tabularx}{0.9\linewidth}{lll}
         \textbf{Roles} & \textbf{Job Titles} & \textbf{Count} \\
         \toprule
         PM & Product manager; Business program manager, etc & 4\\
         Advisor & Chief of staff  & 1 \\
         Software engineer  & Senior software engineer  & 1 \\
         Researcher  & Applied scientist & 2 \\ \hline
    \end{tabularx}
    \caption{Distribution of roles in our interview participants}
    \label{tab:InterviewRoles}
\end{table}

\subsection{Interview protocol}
\label{coding_interview}
Interviews were conducted remotely over video calls by the research team with an average duration of approximately 45 minutes per interview. The first half of the contextual inquiry was intended to gather general information on participants' job roles, working practices and their perceptions on their schedules. We asked open-ended questions such as ``Can you tell us a little bit about your job function and what role meetings play in your day to day work?'' and ``What do you think makes a good schedule for you, and why?'' 

The second half of the contextual inquiry sought to develop a better understanding of what an ideal schedule meant to different participants, what criteria they use to describe an ideal schedule, and what kinds of changes they would make to their schedule if they had no constraints to align their schedule with their preferences. We also elicited potential factors that might influence participants' preferences and practices. 

To accomplish this, the second half of the contextual inquiry grounded participants' responses in the context of their daily activities and needs by focusing on an instance of a “typical week” in the participant’s calendar and  questions about the week. 
To do so, we asked participants to select a recent, typical week for them and share their screen with us to be able to walk us through their schedule as they responded to our questions. 
We defined a typical week as a week with a typical load and variety of meetings that didn’t include activities such as travel, special events, etc.  
Furthermore, we asked our participants to walk us through their day-to-day schedule and share how they would re-arrange meetings if given full control, to better align them with their preferences. 
Questions included: 
\begin{itemize}
    \item Would you consider this week an ideally scheduled week? Please walk us through what is desirable about this schedule and what is not.
    \item{Within this week pick a meeting you organized. Is it aligned with your preferences? Why or why not? }
    \item {Within this week pick a day with a typical meeting load. Please describe if you think this day schedule is ideal for you and whether you would re\-arrange the meetings over this day to make it better aligned with an ideal schedule.}
    item {Imagine you have no scheduling constraints and a lot of flexibility; how would you re\-arrange your calendar to make it better aligned with your preferences? }
\end{itemize}

\subsection{Interview analysis}
Interviews were all transcribed using an auto-transcription service and analyzed using affinity diagramming. 
Following the grounded theory approach by Corbin and Strauss \cite{corbin1990grounded},  we separate our coding into an initial open coding and subsequent axial coding on the main concepts and themes. 
Through multiple iterations along with periodic discussions, two researchers identify the main concepts and themes around an ideal arrangement of events on participants’ calendars.

\subsection{Interview results: a generic taxonomy of temporal preferences}\label{section:formative_results}

\begin{table}[t]
\begin{center}
\small
\begin{tabularx}{\linewidth}{>{\hsize=.18\hsize}X | >{\hsize=1\hsize}X >{\hsize=1\hsize}X}
  & \textbf{Cyclical Preferences} & \textbf{Relational Preferences}\\
  \toprule 
   \textbf{Daily} & \textbf{time of day (TOD)}: \textit{specific preferred hour; no meetings at edge of day}  &  \textbf{daily dispersion}: \textit{breaks every hour, only four back-to-back meetings.}\\ 
   \textbf{Weekly} & \textbf{day of week (DOW)}: \textit{preferred days for meetings; light Mondays and Fridays} & \textbf{weekly dispersion}: \textit{spread out meetings evenly across days, cluster meetings on a few days} \\ 
\end{tabularx}
\end{center}
\caption{Two preference categories across the daily and weekly time scale. The cyclical and relational preference categories are used to structure the analyses of the results in this paper.}
\label{tab:results_structure}
\end{table}

Throughout our analyses, we found that a high-level taxonomy of temporal preferences emerged.
Specifically, the stated preferences resulted in two categories -- \textit{cyclical} and \textit{relational} -- on different time scales. 
First, subjects described some preferences that repeated chronologically. 
For example, several subjects wished to avoid group meetings specifically on Fridays or avoid meetings during their lunch hour. 
Second, responses also emphasized the relation between meetings in their calendars, such as gaps in between and clusters of events, that depended less on their absolute temporal placement. 
Across both categories, we also observe that they apply at various time scales, be it across hours of the day or days of the week. 
We list examples of these categories in Table \ref{tab:results_structure}, and further describe them in detail below.

\xhdr{Cyclical preferences}
During our exploratory interviews, we asked our participants to walk us through their 
typical day-to-day and weekly work schedules
and share how they would re-arrange meetings if given full control, to better align them with their preferences. 

We observed variation in the \textit{time of day} that seems suitable to participate in meetings, leading to two emergent patterns.
First, avoiding the edge of day and pushing more meetings to the middle of the day seemed like a common theme.  
Participants mentioned this preference both in terms of meeting efficacy and in context of offline obligations.

\begin{quote}
    ``\textit{I'm gonna be more engaged in a meeting in the middle of the day. And towards the end of the day, I'd love to be able to use that time to wrap up.}'' [P8]\\
    ``\textit{so essentially like I'm constrained in the beginning of the end of the day. So I've gotta bring my daughter to daycare, and I've gotta manage her at the end of the day. And so that's like, really, it's kind of interesting. Like, I just have to work within those bounds, which is a wonderful work life balance practice. So anyway, so that's what I'm looking at is the space between taking care of my daughter.}'' [P4]\\
\end{quote}
Second, certain times of day (e.g. morning or afternoon) were deemed suitable for taking meetings by other participants. 
For instance, P1 noted: ``\textit{I really prefer working in the morning, which means more meeting in the morning and less in the afternoon}''.
Both the avoidance of edges, which appeared to be fairly universal, and preference for specific hours, which appeared to be highly personal, are cyclic in that they generalize across different days.

Further, we found that preferences
cycle not only at the day level as times of the day, but also at the week level as \textit{days of the week}.
Past work by Mark et al. \cite{mark2014bored} has also found feelings of boredom and focus may vary depending on the day of the week and in our inquiry we were interested to see how this may impact individuals preferences over arrangement of meetings over different days of week. 
For example, we learned that 5 out of 8 interview participants preferred the edge of the week (particularly Monday or Friday) to be meeting free or have a light meeting load. 
They described a variety of reasons for this preference as they'd like to use Mondays for focused or individual work [P1, P6], to plan the upcoming week [P2, P4] or just taking it slow: ``\textit{I like Mondays to be slow, but useful to have some meetings or get some tasks done.}" [P2].
We also noted some participants avoid scheduling meetings on Monday in anticipation of these meetings being unproductive for the participants: ``\textit{No talk series, no huge events on Mondays. Nothing that needs a lot of brainpower after the weekend.}'' [P3].

Participants also expressed that they avoid big group meetings on Fridays but can do 1-1 meetings, brainstorming sessions, or coffee chats with collaborators. 
P4 explained \textit{``So a lot of the people that I work with like to have no meeting Fridays and I support that. I would like not to have status update meetings on Fridays ... If we're gonna do a research discussion on a Friday that's great. It's nice to sort of reserve for blue Sky experimental interesting stuff.''}
Participants reflected that they reserve certain days to accommodate other team collaborators, especially cross-time zone collaborators. P7 and P8 explained that they worked in North America, but they took Tuesday and Thursday as flexible meeting days to connect with team members from India or Europe.

Thus, we observe from our contextual inquiry that two cyclic patterns of temporal preferences emerge across different time scales.
For one, there is a general avoidance of edges.
At the day level, people find very early or late hours to be sub-optimal for holding meetings; at the week level, Mondays and Fridays are also considered to be least suitable for meetings.
However, there are also highly-personal, context-sensitive preferences for specific times like certain hours of the day, or certain days of the week.

\xhdr{Relational preferences}
While reviewing the responses to how individuals would like to distribute their
meetings across different hours of the day, we also noticed that participants highlighted relational preferences for meeting arrangement that drew upon the relationships between pairs or groups of meetings.

People described concerns with the placement of meetings in relation to their surroundings and how spread out the meetings are placed on the calendar.  
For example, some workdays may consist of back-to-back morning meetings with free 
afternoons devoted to work-related tasks, while other workdays may have dispersed meetings with short gaps in between these meetings. 
This sequence or pattern of meetings throughout a day can be thought of as \textit{daily meeting dispersion}, where a low dispersion rate can result in long blocks of back to back meetings, whereas, highly dispersed meetings will result in spacing out meetings over the day leaving shorter meeting-free time intervals in between them.
\begin{quote}
   \textit{``If I look at my calendar and I'm back to back all day, that's really worrying and so a sense of balance across the schedule is also really important. ... if I have back to back meetings for more than two hours that I need to engage in, I fall behind on other tasks during the day because I don't have breaks between meetings.'} [P1] 
\end{quote}

Similar to daily dispersion preferences, \textit{weekly dispersion preferences} also appear as individuals tend to either group meetings in close temporal proximity (e.g. within the same day) or spread them out as much as possible (e.g. across days). 
In comparison to daily dispersion, for weekly dispersion interview participants exhibited a noticeable tendency to distribute their weekly meeting vs. non-meeting workload across multiple days to create rhythms that aligned with their preferences. Interestingly, we also observed a tendency to cluster meetings of certain type into the same day to reduce context switching between meetings (e.g. [P3]), maximize opportunities for in-person collaborations (e.g. [P4]) or to accommodate cross-timezone collaborators (e.g. [P6]).
\begin{quote}
    \textit{So generally I try and cluster (team) meetings on Wednesdays. This is partially because my team comes to the office on Wednesdays, or at least we try to come into the office on Wednesdays so that we can have some overlap and so most of the meetings that I have on Wednesday are with my team.} [P4]
\end{quote}
In unpacking [P4] quote above we note the interaction between a relational arrangement preference (grouping team-meetings on the same day) and a cyclical preference (dedicate Wednesdays to working from the office).

\xhdr{Summary}
Thus, the two main classes of preferences emerged from our contextual interviews can be structured as a 2 (cyclical or relational) $\times$ $n$ (time scales; we consider $n=2$ for hours of the day and days of the week) grid as shown in Table \ref{tab:results_structure}. 
In this table the first column represents \textit{cyclical} preferences that repeat chronologically. As illustrated by our examples, these preferences were most salient at time of day granularity (upper left cell), and day of week granularity (lower left cell). 
Thus the rows of
the table indicate chronological granularity, either at the day level (top row) or week level (bottom row).
The second class of meeting arrangement preferences, the \textit{relational} preference class, is represented by the right column in our table and can also occur at the daily or weekly level.

%% file: 4-Method.tex
\section{Main Study: Methods}
In order to address our research questions, we used a mixed-methods study comprised of three main steps within a large, multinational technology corporation\footnote{Anonymized for review.}.
To address \textbf{RQ1-Preferences}, informed by the insights from the formative contextual inquiry  we developed and distributed a survey within the large technology corporation. The primary goal of this survey was to validate the preference categories and reasoning of workers in a broader audience.  
Secondly, to answer \textbf{RQ2-Practices}, we perform a large-scale trace analysis of millions of meetings conducted at the technology corporation  during the month of October 2022. 
Finally, to further investigate the alignment between workers stated preferences and their observed scheduling practices, we linked telemetry and survey responses for 101 consented survey respondents and further conducted analysis on each dataset. We describe each of these steps below.

\subsection{Mixed-Design Survey}
We conducted a survey asking employees about their perceptions of their work schedules and elicit their preferences. The qualitative coding of the interviews described in the previous section helped us identify key research questions and themes to focus the survey and inform the question design. This survey was conducted in the context of a larger research project on how people arrange their calendars, and it included multiple sections on topics such as hybrid meetings, schedule fragmentation, calendar privacy and disclosure, and tool design.

\subsubsection{Survey design and protocol}
This study focuses on three survey components relevant to our research questions. 
First, the survey included an introductory section that presents participants with an information sheet and asks participants for their consent.
The front matter of our survey is attached in Appendix~\ref{appendix:consent}.
It also solicited background information such as participants' job function, timezone, number of people supervised, and department.

Second, The survey included a segment on daily scheduling preferences asking participants to identify whether each one hour block from 7am to 9pm (or before 7am or after 9am) was a) a typical work hour, b) a preferred work hour, or c) a preferred meeting hour. 
This is intended to elicit stated, explicit perceptions of how employees' work is distributed across the day, and whether or not these are aligned with their preferred work cadence. 
To rationalize these responses, respondents were also asked an open-ended question about their preferences for meetings across the day: ``\textit{Please elaborate why these meeting hours are ideal for you.}''. This segment ended with additional questions about an ideal placement of meetings over their daily schedule probing the dispersion of these meetings.

Third, we also explored preferences around weekly arrangement of meetings by asking participants to describe aspects of one week in their calendar that is ideally scheduled and any changes they would make to better align their weekly schedule with their preferences.
The specific questions associated with these are shown in Table \ref{tab:all_questions}.

\begin{table}[t]
    \begin{tabularx}{\linewidth}{>{\hsize=.15\hsize}X >{\hsize=.15\hsize}X >{\hsize=.75\hsize}X}
     \textbf{Label} & \textbf{Type} &\textbf{Question} \\
    \toprule
    \multicolumn{3}{l}{\textbf{\small{Background Information}}} \\
     \celltext{B-JOB} & \celltext{Text} &\celltextpl{{What is your work role / title? (e.g. Senior Software Engineer)}} \\
     \celltext{B-TZ} & \celltext{Single Choice} & \celltextpl{{What is your time zone?}} \\
     \celltext{B-SUP} & \celltext{Text} & \celltextpl{{How many people do you supervise directly?}} \\
    \celltext{B-DEP} & \celltext{Single Choice} & \celltextpl{{Which department do you currently work under?}} \\
    \midrule
    \multicolumn{3}{l}{\textbf{\small{Cyclical Preferences: Day}}} \\
     \celltext{TOD1} & \celltext{Multiple Choice} & \celltextpl{{Which of the following hours are typically part of your working hours?}} \\
     \celltext{TOD2} & \celltext{Multiple Choice} & \celltextpl{{If you had full control over your schedule, how would you distribute your work within a day?}} \\
     \celltext{TOD3} & \celltext{Multiple Choice} & \celltextpl{{What are your preferred times of day to participate in meetings?}} \\
     \celltext{TOD-OPEN} & \celltext{Text} & \celltextpl{Please elaborate why these meeting hours are ideal for you.}\\
    \midrule
    \multicolumn{3}{l}{\textbf{\small{Relational Preferences: Day}}} \\
    \celltext{DD1} & \celltext{Multiple Choice} & \celltextpl{{On a typical day with a typical quantity and variety of meetings, which of the following describes an ideal placement of meetings?}} \\
    \celltext{DD2} & \celltext{Multiple Choice} & \celltextpl{{What would be your preferred duration limit for back-to-back meetings?}} \\
    \celltext{DD-OPEN} & \celltext{Text} & \celltextpl{Please briefly elaborate why this placement of meetings over a day is ideal for you?} \\
    \midrule
    \multicolumn{3}{l}{\textbf{\small{Cyclical Preferences: Week}}} \\
     \celltext{DOW-OPEN1} & \celltext{Text} & \celltextpl{Please describe which aspects of this week schedule is ideally scheduled and which aspects are not?}\\
     \celltext{DOW-OPEN2} & \celltext{Text} & \celltextpl{{If you had the freedom, would you make any changes to the arrangement of meetings over the week to make it closer to your ideal schedule? What would they be?}} \\
    \midrule
    \multicolumn{3}{l}{\textbf{\small{Relational Preferences: Week}}} \\
     \celltext{WD} & \celltext{Multiple Choice} & \celltextpl{{Imagine you need to schedule three meetings during a week and have full control over when they take place. How would you prefer to arrange these meetings?}} \\
     \celltext{WD-OPEN} & \celltext{Text} & \celltextpl{Please describe the main considerations you had while scheduling these three meetings.}\\
    \bottomrule
    \end{tabularx}
\caption{Summary of survey questions and type of response elicited. For TOD1 through TOD3, participants select all one hour blocks between 7am and 9pm that apply for the question. For daily and weekly dispersion questions (DD and WD), participants select one option out of various specified meetings placement over daily and weekly schedules.} 
\label{tab:all_questions}
\vspace{-5mm}
\end{table}

\subsubsection{Survey data collection}
In total, we obtained $n=180$ responses to the survey, of which $n=165$ qualified as eligible employees (we removed those in sensitive or identifiable positions, such as C-suite level workers). Among these eligible participants, 94\% said they had recently worked across time zones. Participants were also equally split between working fully remote (49\%) and hybrid (49\%) with only 2\% working from the office on all days.
Respondents' ages ranged from 18 to 55 years (M = 41.8, SD = 9.1), with 37\% identifying as female and 61\% identifying as male.
Respondents came from a diverse set of roles ranging from product managers and software developers to consultants and administrative assistants; the majority had been working at the
company for more than 6 years (52\%). 
Multi-tasking on different projects was a common trend among them, with 74\% working on at least 3 projects in parallel at the time of completing the survey. 

\subsubsection{Open-ended question analysis}\label{survey-coding}
In addition to analyzing multiple-choice questions like TOD1 quantitatively, we also conducted qualitative analysis on the open-ended questions in the survey by following a similar grounded theory approach with two researchers as in Section~\ref{coding_interview}.
In terms of cyclic preferences, the two coders conducted a cross-examination of the responses to three multiple-choice question (TOD1, TOD2, and TOD3) alongside the qualitative TOD-OPEN. 
From this, we were able to elicit the preferred segment of the day that is ideal for taking meetings for each participant, how that segment is placed within their specified work hours, and their rationale for this preferred placement of meetings over their workday. 
Coded themes result in five primary arrangement categories representing how participants preferred meeting block fits within their (work)day. 
To categorize workers' preferences on weekly meeting arrangement preferences, the same researchers coded and categorized weekly meeting arrangement preferences by grounding responses on an instance of a typical week in their calendar (DOW-OPEN1 and DOW-OPEN2).

A similar procedure with two coders is used to understand relational preferences.
In order to better understand ideal patterns of spacing out meetings over a workday we analyzed and coded the rationales behind the preferences elicited in question DD1, DD2, and DD-OPEN and specify themes on daily dispersion preferences.
This was done analogously to the procedures for the TOD questions, described above.
We further surfaced differences in weekly dispersion preferences by asking participants how they would ideally choose to distribute three new meetings over the week (question WD).
The accompanying rationales articulated in WD-OPEN were again qualitatively coded by the same two researchers.
The coding results in four weekly dispersion preference patterns.

\subsection{Large Scale Telemetry Analysis}

Our quantitative results are primarily derived from an \textit{in situ} telemetric trace~--~i.e. recorded behavioral data~--~of meetings at the company during the month of October 2022 (henceforth the \texttt{Large-scale Telemetry Dataset}) 
We processed a sample of all events scheduled into employees’ work calendars in a major commercial Web email client, the company’s primary information management system.
Each data point includes information about when the event was scheduled, its scheduled duration, the individuals who attended, and the individual who organized the event. 
We additionally restricted events to those recorded as meetings (as opposed to appointments or ``out of office'' blocks). 
Finally, we restrict our dataset to  meetings a user joins online. This reflects interview subjects indicated that their calendar not a ``true representation'' of their time, and they plan not to attend many meetings that appear on their calendar.
While respondents indicated that the majority of attended meetings were online during the analysis period, 
we do acknowledge that this approach under-counts true meeting load throughout our analysis.
Our resulting dataset contains 6.9 million meetings between 211 thousand individuals.

We additionally consider two coarse-grained pieces of information from this dataset.
First, we inferred attendee timezones in meetings by parsing geolocation metadata and mapping them to offsets from Coordinated Universal Time (UTC), adjusted for daylight savings based on when events occur. 
Second, we derive generic job titles by following corporate guidelines about job functions.
At the organization we studied, this included Software Engineering, Human Resources (HR), Science, Product Management (PM), Sales and Marketing, Design, and Legal functions. 
To protect employee privacy, all potentially sensitive information like individual job titles and cities was then removed after we computed coarser-grained tags like managerial level and time zones. 
At no point did our data contain any explicit identifying information like names and emails.
Data access was also restricted only to members of the research team with research ethics training.

\subsection{Survey-Linked Telemetry Data}
In addition to analyzing the survey and telemetry separately to address our research questions, we also perform a small-scale, exploratory analysis of the linked calendars of consenting survey respondents (henceforth the \texttt{Survey-Linked Telemetry Dataset}).
We asked respondents if we could link their survey responses to their telemetry data for this part of the study, to which
$n=101$ of 165 (61\%) people consented and were therefore linked. 
Although the linking procedure required email, this information was immediately discarded after survey responses and telemetry were joined. 
No identifying information was used in subsequent analyses.

\subsection{Ethical Considerations}

Because our work analyzes employees' perceptions, through our interviews and surveys, and the practices, through our telemetry analysis, we are aware of potential concerns over the protection of individuals in our study~\cite{ball2010workplace}.
In addition to the aforementioned steps taken to protect employees, we further address this from several directions.
First, our work has undergone a formal privacy review at the corporation to ensure individual employees are not at risk of identification. 
Second, our interview and survey participants provided consent for the use of their responses in our research shareouts including this paper. We have additionally obtained the survey participants consent to link their responses to their telemetry data. We are providing the consent form that is used for our study in the Appendix.

Third, all results concerning worker behaviors derived from the telemetry were pseudonymized, with no names or email addresses. We  also ensured, this data was obtained and presented at a coarse-grained, aggregated level; and behavioral data from individual employees is neither analyzed nor shown in this study.
We further discard potentially sensitive metadata, such as city-level geolocations, user demographic information and specific job titles indicating seniority alongside job function. 

Fourth, we explicitly avoid quantifying performance indicators like worker productivity that raise concerns about surveillance.
Instead, our paper is concerned with aiding stakeholders~--~managers, policy-makers, and individual employees themselves~--~find ways to accommodate the diverse temporal preferences held by their colleagues.
In the context of promoting healthier distributed teams, our data-driven methods aim to help workers from a pro-social perspective~\cite{nissenbaum2004privacy}.
Finally, access to the dataset was restricted to the research team involved in this project. We provided details around what data we collect, how we use and store the data in the consent form provided in the appendix.

%% file: 5.1-Results-Preferences.tex
\section{Main Study Results: RQ1 - Temporal Preferences}\label{section:preferences}

Having described our methods and how they are derived from the formative study in Section~\ref{section:formative}, we now present the results of our analyses.
We structure our findings into two sections.
Section~\ref{section:preferences} addresses \textbf{RQ1} by structuring survey responses according to the taxonomy of temporal preferences surfaced by our formative study. 
To address \textbf{RQ2}, Section~\ref{section:practices} then evaluates whether workers' actual calendars are structured according to these temporal preferences by investigating the telemetry data we collected.

\subsection{Cyclical, Day-Level: Times of the Day}

Throughout  our formative study (see Section \ref{section:formative}) we identified cyclical preferences which represent meeting times that repeat chronologically, and on various time scales.
To what extent are these temporal preferences, as categorized in Table~\ref{tab:results_structure}, reflected in the responses obtained from our more structured survey questions?

\begin{figure}[h]
\centering
\minipage[t]{0.80\textwidth}
    \vspace{0pt}
    \includegraphics[width=1\linewidth]{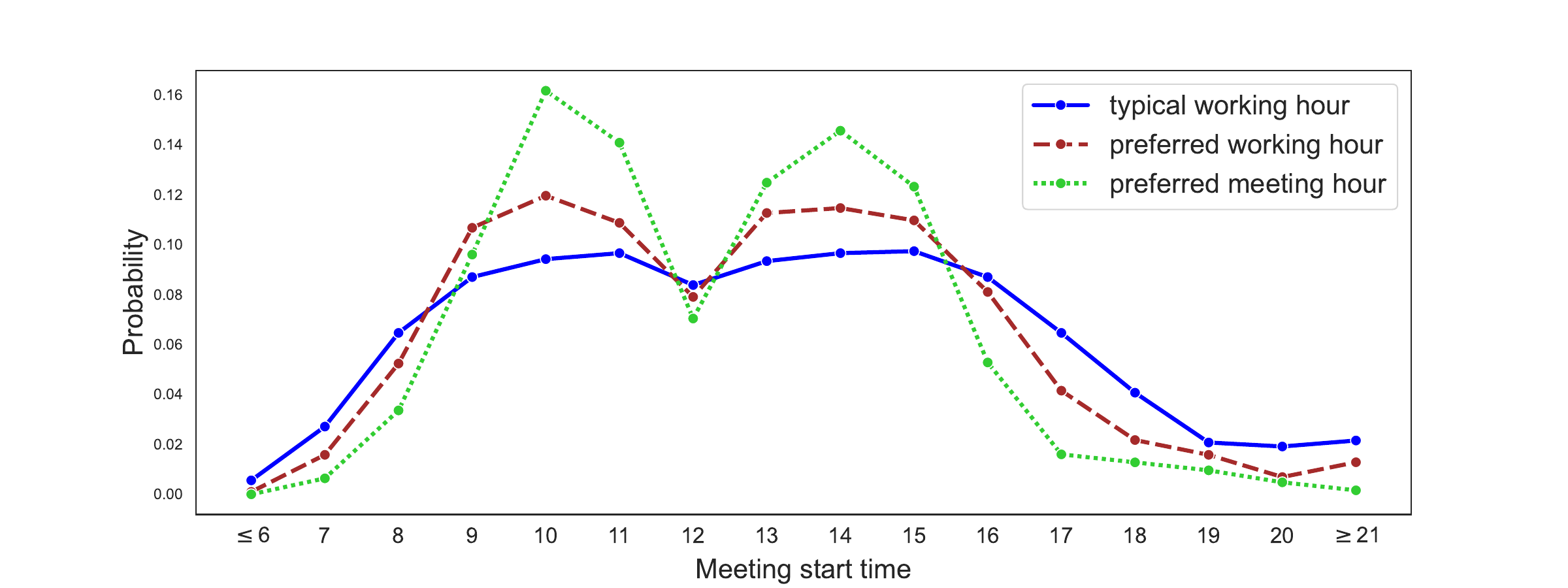}
\endminipage\hfill
\caption{
    The probability that a given hour of the day is selected by participants in our survey as representative of their typical working hours, their preferred working hours, and their preferred hours for holding meetings.
}
\label{fig:stated_prefs}
\vspace{-1em}
\end{figure}

We first consider cyclical preferences at day-level by characterizing how self-reported daily rhythms are structured across workers' preferred meeting hours. 
These are then contrasted against their typical and ideal hours for personal work.
In order to determine preferred meeting and working hours in our survey, we total the number of times each hourly block was selected by participants for the TOD (1-3) survey question in Table~\ref{tab:all_questions}. 
We then normalize this count by the total number of hour blocks selected by all participants for that question.
This yields the probability that each hour block is viewed as being reflective of workers' actual work hours (blue line), their preferred work hours (brown line), and their preferred meeting hours (green line), as shown in Figure~\ref{fig:stated_prefs}.

At the daily time scale, we find that people display clear preferences for scheduling meetings towards the core of the ``traditional'' workday.
Only 3.5\% of hours before 8am and after 6pm are considered preferable (green line), with preferable meeting times being $>$95\% likely to fall within the core of the day.
Our participants also generally avoid meeting at 12pm for lunch (which is reflected in their open-ended responses) and there was a slight preference for morning meetings over afternoon meetings.
In comparison, respondents reported more flexibility with preferred times for non-meeting work (brown line; 7.4\% outside of 8am-6pm).
Similarly, typical working hour responses include every hour between 6 am and midnight (blue line), with 13.5\% of typical hours outside of 8am-6pm. 
Together, these observations are concordant with our preliminary findings in Section~\ref{section:formative}, and suggest a generally-held preference for scheduling meetings towards the middle of the day and away from the edges.

Nonetheless, this information alone fails to capture \textit{why} workers hold these preferences. 
We therefore synthesize high-level patterns between user responses to the three hourly TOD questions and their corresponding open-ended rationales, as described by our methodology in Section~\ref{survey-coding}.
Table~\ref{tab:TOD preference} presents each category, accompanied by a description and a representative participant quote belonging to this category.
For each preference category, researchers conducted qualitative coding on reasons to gain insights into their motivations and examine the rationales behind their preferred meeting hours.

\begin{table}[t]
    \centering
    \
    \footnotesize
     \begin{tabularx}{\linewidth}{>{\hsize=.4\hsize}X >{\hsize=1\hsize}X >{\hsize=1\hsize}X >{\hsize=.1\hsize}X}
    \textbf{Categories}&\textbf{Description} &\textbf{Sample Participant Quote} & \% \\ \toprule
    \textbf{Morning} & Individuals who prefer to place all their meetings in their morning block.&  ``I like to meet in the morning, coordinate on actions, and then work in the afternoon to close on items.'' [P21] & 20.6 \\ 
    \midrule
    \textbf{Afternoon} & Individuals who prefer to place all their meetings in their afternoon block. & ``I do my best thinking in the morning, so I'd prefer to use that for focus time outside of meetings.'' [P81] &12.1 \\ 
    \midrule
    \textbf{Middle of the day [MOD]} &Individuals who prefer to place all their meetings in the middle of the day with long uninterrupted time blocks at their edge of day. & ``Middle of the day break from actual work and best for most US time zones.'' [P42] & 15.8\\ 
    \midrule
    \textbf{No meeting at the edge of the day [NMEOD]} &Individuals who prefer not to start or end their day with meetings, while meetings can be scheduled at any other time of day. & ``I like the end and beginning of my day free for email and planning'' [P15] &35.8\\ 
    \midrule
    \textbf{All day [AD]} & Individuals who don't have a strong preference for their meeting hours as long as they land within their working hours. & ``I am available for meetings all the time I am at work. I prefer not to meet during my lunch hour. However, since I work with people in different time zones, I accommodate for that.'' [P13] & 19.9 \\ 
    \bottomrule
    \end{tabularx}
    \caption{Description of time of day meeting preferences, sample quotes, and percentage of responses falling within each category.
    }
    \label{tab:TOD preference}
\end{table}

Overall, 80.1\% of our participants perceive certain times of day as more suitable for engaging in synchronous collaborative work facilitated through meetings. 
Conversely, only 19.9\% of our participants stated that meetings can occur at any time of day as long as they are within their working hours. 
This observation highlights the distinction between working hours and preferred times to hold meetings. 
Responses from the 80.1\% with distinct preferences can be further delineated into the two preference subcategories described in Section~\ref{section:formative_results}: general avoidance of edges and personal inclinations for specific hours of the day.

In terms of temporal edges, we found that more than 50\% of our participants prefer to hold meetings in the middle of their day (16\%) or avoid the edges of the day (35\%). 
This aligns with previous research\cite{tang2011your} where participants explained that they may have conflicting personal or family obligations during the edges of the day, such as commute time or other routines.
Furthermore, some participants expressed a desire to use the beginning of their day as a ``warm-up'' period, during which they catch up on emails and asynchronous communications or plan their day ahead. 
They also mentioned the importance of wrapping up pending action items or follow-ups towards the end of the day when their memory is sharpest for recalling actionable items.

In terms of specific hours, 33\% mentioned an ideal time of day for holding meetings, with some indicating the morning (21\%) and others preferring the afternoon (12\%). 
However, as suggested by our formative study results, these also appear to be highly personal in the survey responses.
To what extent can underlying factors be observed in the respondents' stated personal preferences for specific hours? 
Below, we find that two types of factors, \textit{chronobiological} and \textit{organizational}, can explain the variance in how individuals hold cyclic preferences for specific hours.

\xhdr{Chronobiological factors}
Qualitative coding on the TOD preferences described in Section \ref{survey-coding} resulted in two main underlying themes representing the motivation behind placing all meetings in the morning block versus an afternoon block.
Past research has shown that circadian rhythms can have a strong influence on sleep, mood, and cognition~\cite{schmidt2007time}. 
The first theme relates to individuals selecting a time of day for collaborative or individual work based on their energy levels and alertness. 
This category of temporal rhythms is influenced by  individuals’ chronobiology, i.e., their inner clocks and diurnal preferences \cite{janbocke2020finding, jun2019circadian, schmidt2007time}. 
The second theme highlights that individuals typically have an awareness of their most productive time of day, and they tend to allocate that period for important work tasks.
This finding is consistent with previous studies that have also recognized developers' inclination to allocate their most productive time to coding and bug-fixing tasks ~\cite{meyer2019today}.
However, there are variations among individuals regarding which type of activity is best suited for their peak time of day. 
For instance, some participants stated that they ``\textit{do my best thinking in the morning, so I’d prefer to use that for focus time outside of meetings.}'' [P81], and that their ``\textit{brain is more alert in the mornings for meetings. Would prefer larger blocks of time in the afternoons to get on with tasks.}'' [P41].

While individual chronobiology influences their preferences for meeting hours, a larger group of participants structure their work in a way to maximize their productivity and work outcomes. 
In our context, for example, is it more productive to start the day with focused or individual work and then proceed with meetings, or the reverse? 
For [P39] taking meetings later in the day is preferred as it ``\textit{allows for prep for the day, tackling big actions, then getting and sharing updates}'', while for [P70] taking the meetings in the morning is preferred because ``\textit{In the afternoon I would like the opportunity to act on the action items that come out of the morning meetings. Also, I'd prefer flexibility in the latter part of the day.}''.
Together, these observations show that

\xhdr{Organizational factors}
Although there is a general consensus among participants regarding the preference for scheduling meetings during traditional work hours, we discovered a diverse range of individual idiosyncrasies when delving into the underlying reasons behind their specific time of day preferences. 
Inspired by the previous CSCW literature ~\cite{olson2000distance, meyer2019today} on factors that impact individuals' work rhythms, we investigated the following user characteristics that could provide insights into the diversity of work rhythm preferences: commute requirements, work modality, nature of the work (hands on type of work, work with tools and machines, work that involves scientific and creative tasks, work that involves collaboration and coordination with others), and meeting load.
To investigate this, we binarized each of these organizational factors (such that, e.g., a person either mentioned commuting or not), and then iteratively conducted corrected Fisher exact tests per factor against the categories in Table~\ref{tab:TOD preference}.

We conduct the test between workers' commuting requirements -- whether or not they stated the need to commute -- and time of day preference categories.  
We find strong statistical significance ($p \leq 0.001 ^{**}$), with steering away from edges of day for meetings (i.e., NMEOD) being relatively more common for individuals who need to commute to work.

We also find that the nature of the work influences workers' preferred times to participate in meetings.
A Fisher's exact test showed statistical significance ($p \leq 0.05^*$), indicating that there is a significant difference in meeting preferences between workers whose daily tasks involve more hands-on work and those who do not. 
Participants engaged in hands-on tasks tend to prefer meeting in the morning, while workers involved in scientific work tend to prefer meeting in the afternoon ($p \leq 0.05 ^*$). 
Collaboration and coordination with other co-workers can largely impact an individual's schedule, especially in the team, since it decides whether the communication happened synchronously and asynchronously. 
Another Fisher's exact test showed statistical significance ($p \leq 0.05 ^*$), suggesting that meeting preferences differ between workers whose daily tasks involve more collaboration and those who do not. 
Participants engaged in collaborative work tend to avoid meeting at the edges of the day. 
These patterns highlight the influence of individual work characteristics, such as commuting requirements and the nature of tasks, on meeting preferences. 
Workers' preferences are shaped by their unique circumstances and the optimal timing for their work activities.

\subsection{Cyclical, Week-Level: Days of the Week}

We have observed how workers hold preferences for and against meeting at different times of the day; however, there is good reason to believe that they also have cyclical preferences at the week level.
Previous research has shown that workers' feelings of boredom and focus can be associated with the day of the week, with higher levels of boredom reported, particularly on Mondays ~\cite{mark2014bored}. 
To gain insights into users' preferences regarding work rhythms at the weekly level, we included specific questions in our survey that focused on eliciting preferences for weekly meeting arrangements.
These questions prompted participants to consider a typical week in their calendar (referred to as DOW-OPEN1-2) and ensured the responses are grounded in that context. 
By examining their preferences in this manner, we aimed to better understand how participants prefer to structure their workdays throughout the week.

We observe a similar set of patterns that are closely aligned with the TOD categories discussed in the previous subsection (refer to the Table \ref{tab:TOD preference}). While \textit{Morning} and \textit{Afternoon} were frequently referenced as participants were asked about their daily meeting arrangement considerations, we see Monday and Friday are commonly referenced when shifting our focus to an ideal weekly schedule. ``\textit{Not an ideal week. * Meetings are spread out throughout the day. I try to have minimal meetings on Fridays and this Friday is packed.}'' [P70]. In addition, participants mentioned that they would like to schedule meetings on specific days (41.2\%) instead of across weeks. One reason is to balance the in-person days and work-from-home days in the hybrid modality. As P170 mentioned:\textit{``If possible, I prefer to have meetings spaced out, but I also prefer to meet in person if possible. So it's better to cluster on one of my in-office days than to spread to WFH days.[P170]''}.  35.7\% of participants prefer to have meetings across the week days. Most common reason people mentioned is to accommodate to other team members, especially the team from other time zones. For instance,  P129 mentioned that across days meetings avoid too many back-to-back meetings \textit{ ``Across different days, so I could plan prep time before each. Could focus on each completely. Assuming they are with people in US TZ - so across days won't mean too much evening work compared to consecutive.''}

\begin{table}[t]
    \centering
    \
    \footnotesize
     \begin{tabularx}{\linewidth}{>{\hsize=.4\hsize}X >{\hsize=1\hsize}X >{\hsize=1\hsize}X >{\hsize=.1\hsize}X}
    \textbf{Categories}&\textbf{Description} &\textbf{Sample Participant Quote} & \% \\ \toprule
    \textbf{AllWeek} & Individuals who prefer to place meetings across the week & I have a lot of other meetings so, spreading them out over several days helps avoid too much clustering of meetings [P22] & 35.7\%\\ 
    \midrule
    \textbf{SpecificDays} &Individuals who prefer to place all their meetings in the specific weekdays, eg Tuesday. & My preferred days for meetings would probably be Mondays and Thursdays, as I set up and check in through meetings [P121] & 41.2\% \\ 
    \textbf{NM-EOW} & Individuals who prefer not to place meetings at the edge of the week -- Monday or Friday &  I don't like to be in the office at 7 am on a Monday, also I prefer a quiet and early end of my day on Friday afternoon,[P67]& 5.5\% \\ 
    \bottomrule
    \end{tabularx}
    \caption{Description of the day of week meeting preferences, sample quotes, and percentage of responses falling within each category.}
    \label{tab:DOW preference}
\end{table}

\xhdr{Factors associated with Day of Week preferences. }

As we observe in the interview, participants reflected that their meeting loads and collaboration with others constrained how to arrange meetings. We conducted one-way ANOVA test between participants' meeting load and day of week preferences. Contrary to expectations, results showed that participants with a higher meeting load showed a significant preference for meeting at specific days rather than spreading them out(F(2,165)=3.23, p <= 0.05 $*$). This finding is intriguing because, in reality, individuals with high meeting loads often find themselves dedicating most of their weekdays to meetings. The discrepancy between the stated preference and the actual practice suggests that participants may have a desire for more balanced schedules but face practical constraints in achieving them due to their meeting load. 

\xhdr{Summary} 
In sum, we can observe a consistent pattern among participants where they do not prefer meetings at the edges of the day, such as early morning or late afternoon. This trend is preserved in a weekly manner as well, where meetings are not preferred on Monday and Friday. In our factor analysis, we identified additional factors that influence workers' preferences for work rhythm, beyond those identified in the previous study on value creation and sentiment~\cite{meyer2019today}. These factors include workers' commute requirements and the nature of their tasks. Commute requirements emerged as an influential factor, with workers who have to commute preferring to schedule meetings away from the edges of the day, likely to accommodate travel time. Furthermore, the nature of tasks performed by workers also played a role in their meeting preferences. Those engaged in hands-on tasks showed a preference for morning meetings, while workers involved in scientific work tended to lean towards afternoon meetings.

\subsection{Relational, Day-Level: Daily Dispersion}
\label{Section:DailyDispersion}
Thus far, we have untangled cyclical preferences for holding meetings at both day-level and week-level resolutions. 
However, our formative interviews in Section~\ref{section:formative_results} also suggest a second class of preferences based on the \textit{relational} spacing between time frames, which can be conceptualized as fragments of work that serve either as efficient debriefings or as disruptions ~\cite{meyer2017work,gonzalez2004constant,epstein2016taking,strongman2000taking}.

In order to gain a better understanding of the ideal patterns of spacing out meetings over a workday, we examined the survey questions DD1, DD2, and DD-OPEN, which are specifically focused on preferences regarding the daily dispersion of meetings. 
These were analyzed according to the methods in Section~\ref{survey-coding}.
Table \ref{tab:daily_dispersion_preference} presents each category along with a description and a representative participant quote that belongs to that category. Additionally, for each preference category, the researchers conducted qualitative coding on the reasons provided by participants. This analysis aimed to gain insights into participants' goals for taking breaks or having consecutive meetings, shedding light on the underlying motivations behind their preferences.

\begin{table}[t]
    \centering
    \footnotesize
     \begin{tabularx}{\linewidth}{>{\hsize=.4\hsize}X >{\hsize=1\hsize}X >{\hsize=1\hsize}X >{\hsize=.1\hsize}X}
    \textbf{Categories} & \textbf{Description}  &\textbf{Sample Participant Quote} & \% \\ 
    \toprule
    NMG & No or Minimal Gaps between meetings to maximize non-meeting block of time in the day for focused work & ``dead time between meetings isn't long enough to actually accomplish meaningful tasks.''[P7] & 25.5 \\ 
    \midrule
    MidG & Middle Ground approach to NMG. There is still a preference to cluster meetings into multiple blocks over the day, but have breaks in between them. & ``I would have blocks of meetings, rather than meetings scheduled throughout the day. Those blocks would have short breaks between the meetings''[P50] & 10.3  \\ 
    \midrule
    BGaps & Gaps are needed to take breaks, refresh, etc. & ``to take a mental plus bio break and recharge myself'' [P95]& 27.8  \\ 
    \midrule
    TGaps & Gaps are needed to take care of task items (quick ones) or action items that emerge from the meetings in a timely manner so they won't pile up. &``Need time to take some important actions from the call or at least add them to tasks.''[P31] & 35.2  \\ 
    \bottomrule
    \end{tabularx}
    \caption{Daily Meeting Dispersion Preference. The table shows daily meeting dispersion preference categories along with corresponding descriptions, sample quotes from survey participants, and the percentages of each category.}
    \label{tab:daily_dispersion_preference}
\end{table}

Overall, 37\% of our participants expressed a preference for clustering their meetings together, resulting in fewer gaps between them (referred to as NMG or MIdG in Table ~\ref{tab:daily_dispersion_preference}). 
For instance, [P51] stated they expressed a preference for having all their meetings packed in the morning, because ``\textit{I can have time to focus (afternoon) to work on items discussed in the meetings, scattered meetings impact my ability to focus.}''.
Interestingly, over half of these participants specified a shorter break time between meetings as their ideal arrangement. 
This finding differs from previous research that suggests developers would like to decrease disruptions and switching time between meetings~\cite{meyer2017work,gonzalez2004constant}. 
We identified two common reasons raised by participants for taking breaks: participants either needed breaks to refresh and recharge (referred to as BGaps), or they used breaks to address quick task items that emerged from the meetings or to attend to preparation tasks before the meeting (referred to as TGaps).  
Participants reflected that interleaving collaborative and individual work benefited from shorter non-meeting blocks before and after meetings. 
For instance, [P42] mentioned that `\textit{`most of my meeting require preparatory work (demo environments, going through the content, etc.) and have tasks I need to follow up with. Therefore I need slack time before and after meetings.}'' [P42]. 

Compared to our results on cyclical preferences, in which we find clear, generally-held inclinations to avoid temporal edges at both the day-level and week-level, relational preferences appear less consistent.
A quarter of respondents want to schedule without gaps, over 60\% prefer gaps, and a tenth need a balance of both. 
This highlights the apparent inconsistencies in the current body of work.
On the one hand, previous studies have highlighted the negative impact of frequent disruptions on developers' workdays, resulting in limited focused time for coding tasks and reduced productivity~\cite{meyer2017work,gonzalez2004constant}. 
On the other, it is also important to recognize the need for breaks during extended periods of continuous work to alleviate fatigue and boredom~\cite{epstein2016taking,strongman2000taking}.

Thus, how can the variance in our survey respondent's perceptions of meeting dispersion be explained?  
After all, understanding workers' preferences for the spacing between meetings allows for schedules that achieve a balance between uninterrupted work periods and necessary breaks. 
This knowledge, in turn, can guide the development of schedules that optimize productivity and well-being for workers, ensuring an effective and satisfying work experience.
In our analyses, we noticed that one organizational factor emerged as a key correlate with preferences for and against clustered meetings: individual workers' overall meeting load.
We delve deeper into this below.

\xhdr{Meeting load and day-level meeting dispersion} Indeed, the heavy meeting load experienced by workers in our study could be a potential reason for the preference for shorter breaks between meetings. 
The need for debriefing time between meetings, especially when facing a high volume of meetings, may drive participants to prioritize utilizing the break time for reflection, thought gathering, or preparation for the next meeting. 
This highlights the importance of considering individual workloads and the specific context in understanding meeting arrangement preferences. 
By minimizing the duration of breaks, participants can optimize their workflow and maintain a sense of continuity for their focus work in their workday.
We further examine the association between meeting load and daily dispersion category.

We conducted one-way ANOVA test between participants' meeting load and daily dispersion preferences. Results showed that participants with a higher meeting load showed a significant preference for clustering meetings on a single day rather than spreading them out (ANOVA, F(2,165) = 5.28, p <= 0.005  $*$). Additionally, they demonstrated a significant preference for back-to-back meetings within a day (referred to as NMG) rather than spreading them out (ANOVA, F(1,165) = 43.0, p < 0.000  $***$). These findings highlight the influence of meeting load on participants' preferences for meeting arrangements. A higher meeting load may lead individuals to prioritize clustering meetings on specific days and having them in close succession within a day.

\subsection{Relational, Week-Level: Weekly Dispersion}
\label{Section:WeeklyDispersion}
Finally, our formative study in Section~\ref{section:formative_results} also suggested that preferences for and against dispersion may extend to the weekly time scale.
Indeed, previous work found that developers sometimes specify some atypical working days as ``meeting day'' or as ``coding day'', which affects how they would like to spread meetings within different days~\cite{meyer2019today}. 
Thus, we further investigate information workers' preferences across different days (i.e., weekly dispersion), or how users prefer to spread meetings across days.

We surfaced differences in weekly dispersion preferences by asking survey respondents about how they would ideally choose to distribute three new meetings over the week (WD).
From closed-ended responses, 50.3\% of participants (N=83) stated that they prefer to split meetings across days, while 40\% of participants (N=66) stated they would distribute these meetings across the same day, either consecutively (22\%) or dispersed (18\%). 
In order to better understand the variation in dispersion preferences we qualitatively coded their rationales articulated in WD-OPEN as outlined in Section~\ref{survey-coding}.
Since this question did not specify the context of the meetings to be scheduled, 21.7\% of our respondents didn't articulate any particular arrangement preference. 
Rather, they mentioned their decision is meeting dependent (13.3\%) or attendee dependent (e.g. accommodating cross-timezone meetings or busy calendars, corresponding to 8.4\% of responses).

While examining these open ended survey responses, we note their rationales closely follow daily dispersion categories previously summarized in Table \ref{tab:daily_dispersion_preference}. 
For example, some participants like P32 prefer to distribute meetings across days to make sure they have sufficient time in between to take care of tasks: ``\textit{Distributing gives me time to process one meeting in my head and use the potential outcome for the consecutive meetings.}''.

\begin{table}[t]
    \centering
    \footnotesize
     \begin{tabularx}{\linewidth}{>{\hsize=.4\hsize}X >{\hsize=1\hsize}X >{\hsize=1\hsize}X >{\hsize=.1\hsize}X}
    \textbf{Categories} & \textbf{Description} &\textbf{Sample Participant Quote} & \% \\ \hline
    NMG-W &Similar to NMG, the motivation is to maximize non-meeting blocks of time for focused or individual work. However, they minimize gaps on certain days of the week to make them meeting days and use other days of the week as focused days. & ``Once I'm in a meetings mindset it's good to continue on with meetings and get them done so I can have a focus day on another day. ''[P24] & 24.6 \\ \hline
    MidG-W &Similar to NMG-DOW, there is a preference to cluster meetings into certain days in the week, but have breaks in between them. & ``breaks in between, but free up other days for work''[P18]& 8.8 \\ \hline
    BGaps-W &Distribute meetings over multiple days to provide more opportunities for taking breaks and not ending up with very packed days. &``I prefer not to feel the pressure of having a lot of meetings in a day. It makes me feel overwhelmed. So, spreading the meeting over 3 different days would be ideal.''[P70] & 19.3 \\ \hline
    TGaps-W &Distribute meetings over multiple days so there is time in between meetings for prep or taking care of tasks that come out of meetings. & ``Scattering them across multiple days gives me a chance to not only work on big priorities but also deal with any emergencies that come up on any given day while not falling overly behind.''[P23] & 47.4\\ \hline
    \end{tabularx}
    \caption{Weekly Meeting Dispersion Preference. The table shows weekly meeting dispersion preference categories along with corresponding descriptions, sample quotes from survey participants, and the percentages of each category.} 
    \label{tab:weekly_dispersion_preference}
\end{table}

Overall, the distribution of weekly dispersion preferences, as shown in Table~\ref{tab:weekly_dispersion_preference} is similar to daily dispersion~\ref{tab:daily_dispersion_preference}.
However, we observed that more participants prefer to distribute meetings over multiple days to allow for focused work on tasks that arise from meetings. Additionally, participants mentioned that their meeting load often limits their arrangement options, forcing them to schedule meetings across multiple days.  As one participant stated, "If I only had 3 meetings and the rest of the week was open, it would be great to schedule the meetings throughout one day, leaving the other days open for focused work. However, with the current meeting load, scheduling three additional meetings this week would mean scheduling them back to back with other meetings and having absolutely no focus time" [P82]. It is important to note that participants acknowledged the presence of uncontrollable factors that can influence their schedule throughout the week, such as the topic and length of the meeting, other people involved, and existing commitments.

\xhdr{Factors for Weekly Dispersion preferences} 
Beyond trying to distribute workload or minimize context switching, individuals who follow a hybrid schedule also have different preferences about the arrangement of meetings over the days they work from the office versus days they work from home. During interviews, some participants (eg. [P4]) sought to arrange meetings with co-workers who work from the office during their in-office days to maximize opportunities for in-person collaborations. We revisited this hypothesis in the survey by capturing participants' agreement with the following statement ``\textit{I prioritize in-person/hybrid meetings to fully virtual ones on days that I'm working from the office}''. 60.7\% of participants either strongly or somewhat agreed with the statement. This reflects that workers distinguish the onsite day versus remote day when they arrange meetings.

To further explore the association between user characteristics and weekly dispersion preferences, we conducted statistical tests. However, we did not find specific associations which indicate that weekly dispersion preferences can be very personal and less universal. 

\xhdr{Summary of Temporal Preferences}\label{section:pref_sum}
In sum, the present Section~\ref{section:preferences} showed that even though some workers stated that they preferred to have meetings consecutively in order to save more time for focus time, most workers stated that they needed breaks between meetings, to debrief themselves or to process tasks in between. Hence, different from previous studies that highlight that disruptions can fragment workers' workday, increase context switching and reduce productivity~\cite{meyer2017work}, most workers valued the breaks. For the dispersion preferences, meeting load is the most influential feature. A higher meeting load may lead workers to prioritize clustering meetings on specific days and save more time for focus tasks.

%% file: 5.2-Results-Practices.tex
\section{Main Study Results: RQ2 - Scheduling Practices}\label{section:practices}
It is not uncommon to find discrepancies between users' preferences and the actual practices they follow. 
Numerous constraints and factors from collaborators, teams, and personal lives often shape the way scheduling decisions are made. 
In this section, our objective is to identify and analyze the misalignments that exist between users' stated preferences and their actual scheduling practices. 
We take two approaches.
Because the vast majority of workers in our telemetric dataset (\texttt{Large-scale Telemetry Dataset}) did not take the survey, we dive into various factors that can influence or constrain scheduling practices by comparing quantitative metrics against preferences in Section~\ref{section:preferences}. 
For survey respondents who consented to telemetric linking, we analyze the linked dataset (\texttt{Survey-Linked Telemetry Dataset}) at a small scale to directly compare preferences and practices. 
To guide our analysis, we again follow the taxonomy illustrated by Table~\ref{tab:results_structure}, with Section~\ref{section:practices_cyclic} analyzing cyclic scheduling behaviors and Section~\ref{section:practices_relational} investigating relational aspects of calendars.

\subsection{Cyclical Scheduling Practices}\label{section:practices_cyclic}
There are good reasons to believe that actual calendaring practices may differ from preferences based on, for instance, how organizations are structured~\cite{meyer2019today,yang2022effects} and where workers are located physically~\cite{olson2000distance,mok2023challenging}.
Thus, we first investigate whether survey respondents who consented to the \texttt{Survey-Linked Telemetry Dataset} exhibit alignment between preferences and practices.
We then explore the association between workers' meeting load, timezones, and job roles with the meeting time that they organized. 

\begin{figure}[ht]
     \centering
     \begin{subfigure}[b]{1\textwidth}
         \centering
        \includegraphics[width=0.8\textwidth]{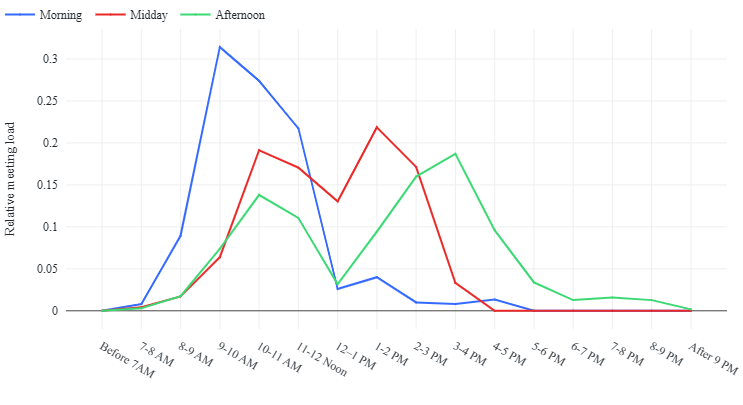}
        \caption{Distribution of survey users' preferred meeting time across a day for three means clusters
        }
        \label{fig:tod_preference_clusters}
     \end{subfigure}
     \vfill
     \begin{subfigure}[b]{1\textwidth}
         \centering
        \includegraphics[width=0.8\textwidth]{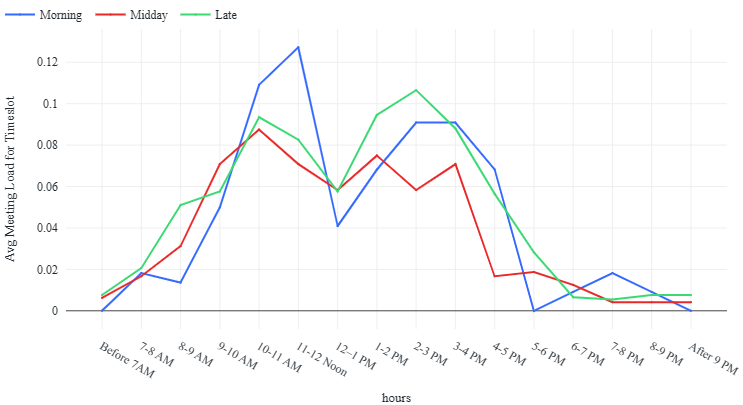}
        \caption{Distribution of survey users' actual meeting time that a user organized across a day for users who are in the same clusters}
        \label{fig:tod_practice_org}
     \end{subfigure}
     \label{fig:feature-novice-expert}
     \caption{Comparison of preferred meeting time and actual meeting time for survey users who are in the same clusters}
\end{figure}

\xhdr{Survey respondents' preferred meeting hours versus their actual meeting hours}
To probe the alignment between participants' meeting practices and preferences, we linked \textbf{survey} respondents telemetry data with survey data to determine whether their actual meeting times reflected their stated preferences (i.e. \texttt{Survey-Linked Telemetry Dataset}). 
We have identified 5 categories of day-level temporal preferences that are explained by a mixture of productivity, biological, and familial motivations. To what extent do these stated preferences align with \textit{actual} work practices? Some groups of users associated with the survey TOD tags in Table \ref{tab:TOD preference} were too small for meaningful analysis. Thus, we took a clustering approach to create coarser TOD definitions.
Using the stated meeting hours a user preferred in the survey, we used K-Means to create three groups of users with distinct TOD preferences. We used the elbow method to find the number of clusters and specify that k=3~\cite{syakur2018integration}. To provide names for each cluster, we used the survey-coded response to correspondingly name these clusters as ``morning'', ``middle of day (MOD)'', and ``afternoon''. Figure \ref{fig:tod_preference_clusters} shows participants' survey preferences within each cluster; they exhibit a clear separation of preferences that aligns with each label: morning, MOD and afternoon.

The distribution of all meetings respondents attended did not clearly reflect their TOD preferences. However, as shown in Figure \ref{fig:tod_practice_org} when we focus on meetings survey respondents organized themselves, we do see an alignment between preferences and practices. The peak for the morning cluster (in blue) occurs before noon, the peak for the late cluster (in green) falls after noon, and the middle of the day cluster (red) is more evenly distributed. We observed that midday workers’ meeting hours in practice shifted towards afternoon time compared with the preferred time.

The weaker alignment between preferences and practices for all meetings when compared to organized meetings could stem
from several different underlying reasons. Organizers could have a greater awareness of their own preferences, or if they are aware, they may be prioritizing their own above other attendees. In line with previous work~\cite{meyer2019today}, open responses highlighted the effects of asymmetries between organizers and attendees on preferences and practices. Power imbalances, such as those between customers and service providers, can exacerbate these effects and make workers lose control on their calendars: ``\textit{The biggest problem is that I don't control my calendar. More than half my day is meetings with customers, and they dictate the meeting time. The rest of my meetings are with internal teams which are scheduled by the Team lead.}'' [P74]. 
For example, time zone differences impose constraints that may not be compatible with individual preferences \textit{My meetings are too spread out across the day. Would prefer to confine to the middle of the day. However, I work with people in the US, Europe, and Asia, so it isn't possible to have times that are convenient for everyone.}'' [P69].

\xhdr{Meeting loads constrained workers' meeting time}
The survey analysis revealed a weak alignment between users' preferences and their actual meeting practices, and meeting load emerged as a key influencing factor. This finding suggests that users with higher meeting loads may face constraints in scheduling meetings, which can result in meetings being scheduled at less preferred times, including the edges of the day.
\begin{figure}[h]
    \centering
    \includegraphics[width=0.5\textwidth]
    {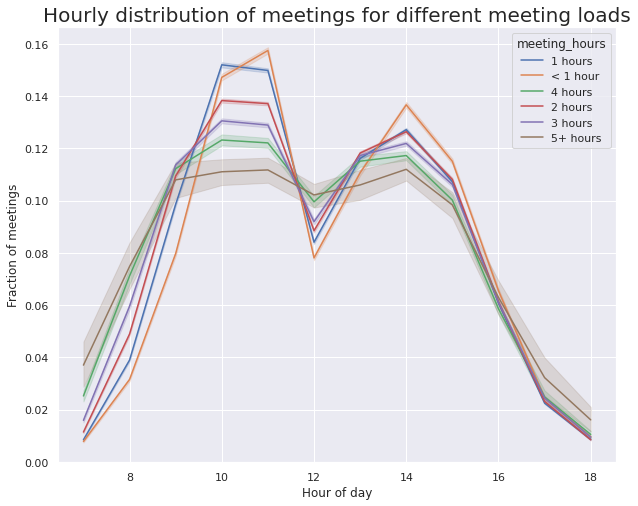}
    \caption{TOD distributions in practice for users with different meeting loads. Meeting loads are grouped by average hours per day of attended meetings. 
    }
    \label{fig:hod_load}
\end{figure}

The telemetry data analysis further supports this observation. 
Figure~\ref{fig:hod_load} shows the in-practice TOD distributions for users in the telemetry dataset with different meeting loads. Users with lower meeting loads, represented by the orange bars (less than 1 hour of meetings daily), exhibit highly peaked distributions of meeting times near the noon hour, aligning with the optimal times identified in the survey. However, as meeting loads increase, represented by the brown bars (5+ hours of meetings daily), the distribution of meetings becomes more evenly spread throughout the day.
Once a user has filled preferred time slots they must also fill less preferred times.
As we see throughout this work, meeting load a dominate in most scheduling practices.

\begin{figure}[h]
    \centering
    \includegraphics[width=0.8\textwidth]{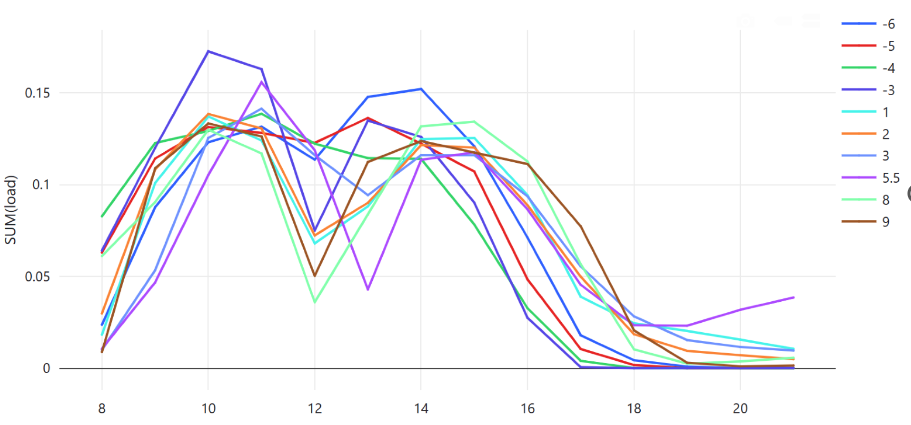}
    \caption{TOD distributions in practice for users in different timezones.}
    \label{fig:tod_xtz}
\end{figure}

\xhdr{Timezones associated with meetings at temporal edges}
The timezone a user works in is also strongly associated with their TOD practices. 
Figure \ref{fig:tod_xtz} shows the relationship between a user's timezone and their daily distribution of meetings on the entire telemetry dataset.  
The graph demonstrates a variance in these distributions based on where employees are temporally located.
For instance, those in time zones ahead of UTC, represented by purple lines, have meeting distributions that are also shifted forwards in their local hour of the day.

\xhdr{User roles limit the time-of-day practices}
Survey analysis revealed that users organizational features and the nature of work included their preferences of meeting time. Here, in the telemetry data analysis, we choose to examine the relationship between workers' job role and their daily meeting time. Job roles can provide insights into the nature of work and organizational features, with software engineers typically engaged in hands-on coding tasks, while managers have different responsibilities. The analysis revealed that job roles, as measured by workers' titles, are indeed associated with time of day (TOD) preferences for meetings.

\begin{figure}[h]
    \centering
    \includegraphics[width=0.8\textwidth]{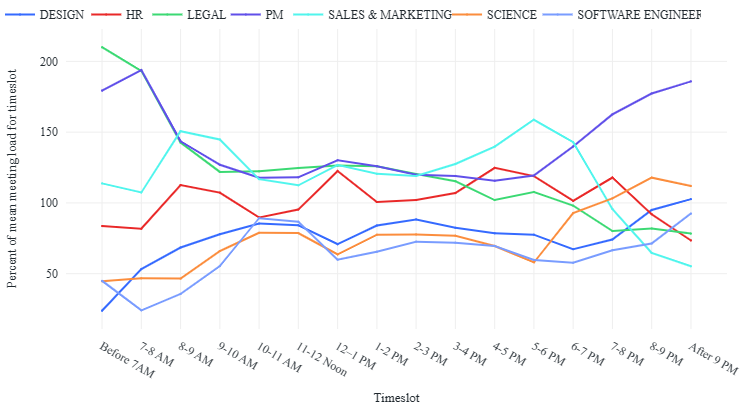}
    \caption{TOD distributions in practice for users with different roles. The Y axis is the percentage change when compared to the average daily meeting load for that hour. }
    \label{fig:tod_role}
\end{figure}

Figure \ref{fig:tod_role} illustrates this relationship by showcasing select job roles and the percentage change in average daily meeting load compared to the overall average user, grouped by different hours of the day. A role that perfectly aligns with the overall TOD distribution would appear as a horizontal line at 1.0.

The analysis suggests that Program Managers (PM) and Legal roles tend to have a higher proportion of meetings at the edges of the day, representing a nearly 200\% increase in average daily meeting load compared to the overall average across all workers. On the other hand, Science and PM roles tend to have a higher proportion of meetings in the afternoon. This indicates that roles associated with more ``hands-on'' work still tend to schedule afternoon meetings, potentially to allow for focused work time in the morning. These findings highlight the influence of job roles on meeting scheduling practices. Understanding these patterns can help organizations better align meeting times with the preferences and work demands of different job roles, optimizing productivity and work satisfaction.

\begin{figure}[t]
    \centering
    \includegraphics[width=0.5\textwidth]{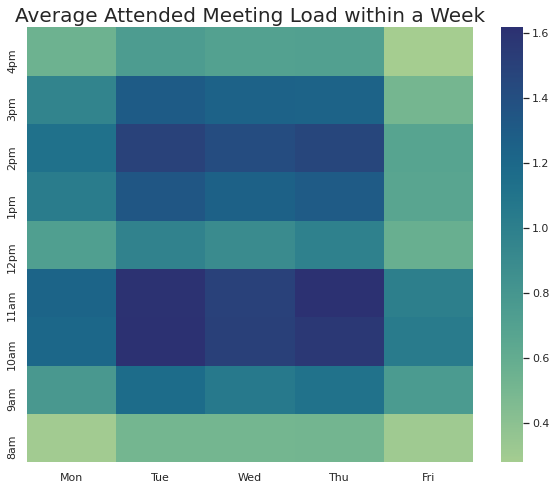}
    \caption{Weekly meeting load for different time slots throughout the week. A value of 1.0 corresponds to the mean meeting load for an hour within the week.} 
    \label{fig:weekly_load}
\end{figure}

\xhdr{Edges of the day versus edges of the week in practice}
In Section~\ref{section:preferences}, we found that time-of-day preferences were only loosely aligned with actual meeting practices. Thus, are day-of-week preferences also similarly misaligned?
Consistent with survey responses and interview responses, edges of the weekdays have lower meeting load than other days. Monday and Friday have the lowest meeting loads (87\% and 63\% of the daily average, respectively), while Tuesday and Thursday are the highest (120\% and 118\%). 
Figure \ref{fig:weekly_load} shows a heatmap that indicates that DOW variations in normalized meeting load are not uniform across all hours. By further zoom in on the specific time across the edge of the week, Monday morning and Friday afternoon is every lower are usually low in their meeting load, even when controlling for overall DOW effects.
For example, Friday afternoon is unusually low in its meeting load when we average the meeting load across different weeks.

\subsection{Relational Scheduling Practices}\label{section:practices_relational} 
Our results in Section~\ref{section:practices_cyclic} illustrate the multitude of scheduling constraints workers face when organizing their calendars according to cyclic preferences, as well as the general prevalence of meetings at temporal edges that are discordant with these preferences.
What behavioral patterns can be observed when it comes to the \textit{relational} spacing of meetings?
In this section, we explore the influence of meeting load and organizational features on workers' relational practices. 
We primarily focus on the telemetry dataset (\texttt{Large-scale Telemetry Dataset}) for this analysis.

\begin{figure}[t]
    \centering
    \includegraphics[width=0.5\textwidth]{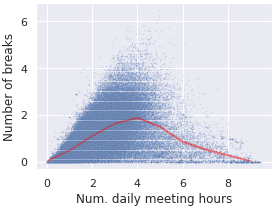}
    \caption{Number of breaks across meeting hours. At a meeting load threshold equal to approximately 50\% of the work day gaps appear to fill in the user’s calendar and they decline.}
    \label{fig:dispersion}
\end{figure}

\xhdr{Meeting load limits users' choices of when to take breaks}
We examined daily dispersion in telemetry at scale. 
To this end, we partitioned the hours between 8am and 6pm into 5-minute timeslots and determined whether a user was free or busy during each time slot. 
We used the \textit{number of breaks} as a measure for dispersion.
For each worker, we split the calendar into 5-minutes blocks and labelled them as either busy or free based on whether an event is scheduled at that time.
Two or more consecutive free blocks, i.e. a gap of ten or more minutes, are then counted as a break.
Figure~\ref{fig:dispersion} plots the distribution of the number of breaks per person-day across the dataset; the red line is the median value on the number of breaks across different numbers of daily meeting hours.

Two patterns are evident from the bell-like distribution.
It appears that people generally schedule more breaks when they have more meeting hours~--~ to 4 hours a day.
This metric suggests that as a user's meeting load rises, a user has a choice for taking breaks, as we see that people took more breaks when their meeting time is longer -- an average 2 breaks between 4 hours of meeting. 

However, after this 4 hour point, which represents 50\% of an 8-hour work day, breaks actually {decrease} with increasing meeting load. 
On the one hand, this could suggest that workers prefer to not have gaps in lengthy days.
Nonetheless, given our observations in Section~\ref{section:preferences}, it seems more likely that this reflects a decrease in \textit{choices} of how to schedule breaks into the day.
In other words, once a person's meeting load reaches to approximately half of the workday, users' choices and variation appear to decline, leading to fewer opportunities for finding breaks.

\begin{figure}[h]
    \centering
    \includegraphics[width=0.6\textwidth]{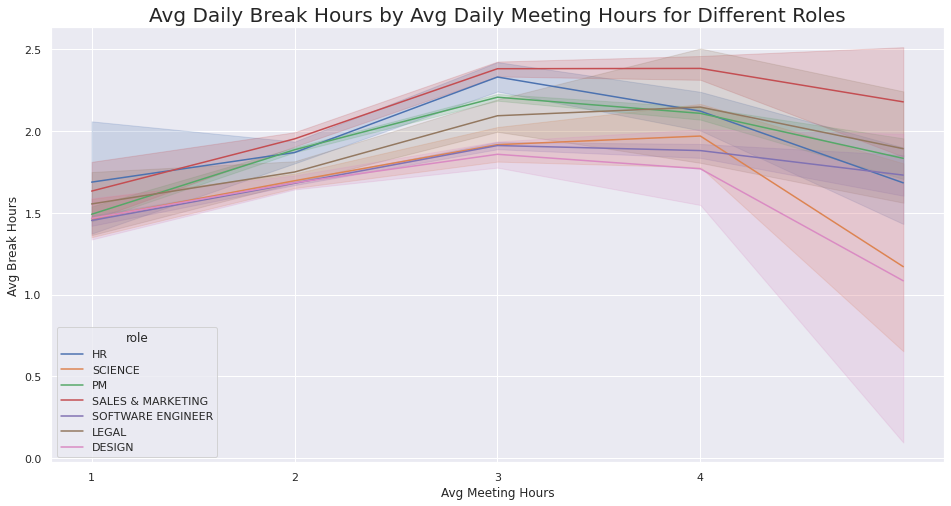}
    \caption{Average number of breaks for different levels of meeting load, broken out by role.}
    \label{fig:break_role}
\end{figure}

\xhdr{Different job roles have a different practice of taking breaks}
We also looked for factors beyond meeting load associated with large-scale telemetry patterns and found that role as
measured by job title played an important effect.
Figure \ref{fig:break_role} shows the average number of breaks for different levels of meeting load, broken out by role.
Human resources and sales have higher break hours, while designers, engineers, and scientists have fewer break hours for the same meeting load.
This hints that a person's perception of the purpose of meetings may correspond to the number of breaks between meetings.
Although this relationship requires further exploration, in roles such as HR and sales, workers seem more likely to view meetings as a place where work is conducted rather than design and science, where work may be viewed as coordinated.

\xhdr{Weekly dispersion is highly personalized}
We measure practices associated with weekly dispersion by calculating a user's weekly distribution of meetings per day and then calculating the entropy of that distribution.
Users with equally distributed meetings will have a uniform meeting distribution of $\frac{1}{5}$ for each day and an entropy value of 2.32 (calculated as $5 \cdot \frac{1}{5} \cdot log_2 \frac{1}{5}$).
Users with all their meetings scheduled on a single day will have a distribution of 1.0 for one and 0.0 for the other four, and an entropy value of 0.0 (calculated as $1 \cdot log_2 1 + 4 \cdot 0 \cdot log_2 0)$.

\begin{figure}[h!]
    \centering
    \includegraphics[width=0.6\textwidth]{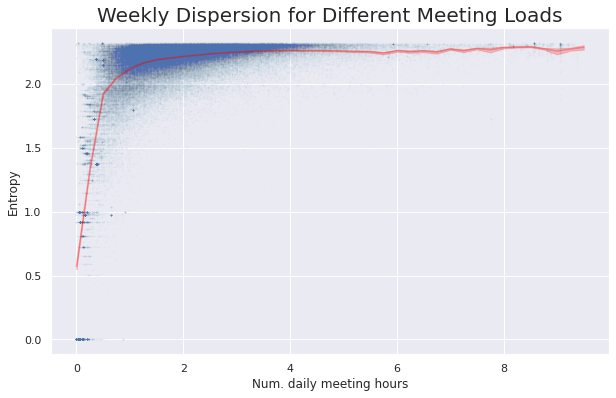}
    \caption{Weekly dispersion, as measured by entropy of the daily distribution of meetings, for different meeting loads.}
    \label{fig:weekly_dispersion}
\end{figure}

As shown in Figure \ref{fig:weekly_dispersion} weekly dispersion, as measured by entropy, varies substantially with meeting load similarly to daily dispersion.
Daily dispersion rises steeply with meeting load before quickly tapering off. 
This may reflect the challenges of packing an increasing number of meetings into a subset of days of the week.
Thus, for meeting loads of at least one hour per day (after the curve flattens), small variations in dispersion values may be meaningful differences.

\begin{figure}[h!]
    \centering
    \includegraphics[width=0.6\textwidth]{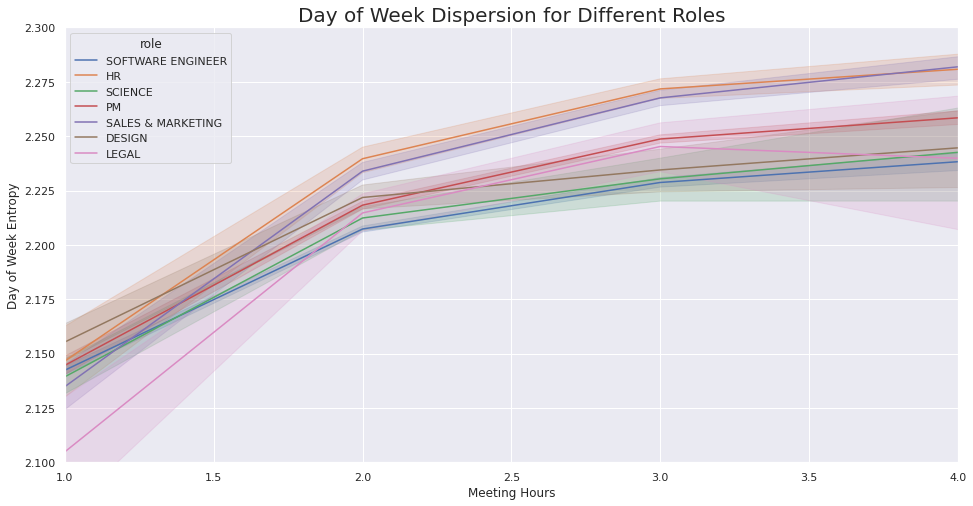}
    \caption{Day of week dispersion, as measured by entropy of the daily distribution of meetings, for different meeting roles and meeting loads.}
    \label{fig:dow_role}
\end{figure}

Figure \ref{fig:dow_role} shows the day-of-week dispersion for different meeting roles and meeting loads. 
As noted in the previous visualization, overall weekly dispersion rises with meeting loads. 
However, we do see patterns in weekly dispersion consistent with those for daily dispersion. 
Once again, human resources and sales have higher weekly dispersion, while designers, engineers, and scientists have lower values, even after controlling for meeting load.
Again, these findings suggest questions about whether dispersion is used to create space for non-meeting time among workers who do not view meetings as their primary vehicle for work.

%% file: 6-Discussion.tex
\section{Discussion}
Prior research finds that people exhibit temporal rhythms in their everyday behaviors which shape and support their work practices as well as their personal lives.
In our mixed-methods characterization of individuals scheduling preferences and practices we noted that as our participants were  reflecting on an ideal arrangement of meetings over their daily or weekly schedule, they focused on some combination of productivity and  well-being. Productivity-focused participants think about an ideal arrangement that maximizes their productivity and enables them to take care of their tasks and action items. The well-being focused individuals, on the other hand, are motivated to create daily and weekly schedules that allow them to take sufficient breaks in between meetings.

Our research on the impact of meetings arrangement over daily and weekly schedules offers new insights for the broader model of how meetings fit into the daily work experience. 
First, we identified two intertwined modes of meetings arrangement that combine cyclical and relational patterns of distributing individual and collaborative work over time.
These two dimensions of meeting arrangement preferences, resulting in four primary preference categories, can facilitate the process of scheduling preference elicitation for new users and contribute to more personalized meeting time recommendations.

\xhdr{Time of Day Preferences}
Our survey results suggest the strong majority of participants (80\%) perceive certain times of day as more suitable for engaging in synchronous collaborative work facilitated through meetings, and that by holding meetings into these ideal times of day information workers can maintain their productivity throughout the day.
While over 95\% of preferred meeting hours fall within the core of day, we noted that employees’ preferred meeting hours are more clustered towards the middle of the day than what they believe their typical meeting hours are. This observation highlights the distinction between preferred working hours and ideal times to participate in synchronous collaboration.
In response to increasing flexibility and diversity of individuals workdays some scheduling tools ask their users to specify their working hours. However, our results argue for a more holistic view of an individual's workday by acknowledging the preferences around meeting hours in addition to their timezone and working hours.

To further probe participants actual meeting behaviors, we examined survey respondents’telemetry to determine whether their actual meeting times reflected their preferences. While we only observed a weak alignment between individuals preferences and practices for all
meetings, once we focused on meeting organizers data in our sample we found their practices are aligned with their stated preferences. 
Organizers of meetings might have a greater understanding of their own preferences and priorities compared to attendees, resulting in fewer scheduling constraints for both the meeting times and organizers themselves, as opposed to non-organizer attendees ~\cite{tullio2002augmenting,berry2007balancing}.
Regardless, this observation highlights the relative authority of meeting organizers over the scheduling process. 
Across our survey and interview responses, a salient reason why existing schedule negotiation patterns may deviate from attendee TOD preferences is that current scheduling systems do not provide workers with sufficient information about their colleagues' preferences.
Thus, preference-aware calendaring tools, could potentially serve as starting points to inform organizers of better and more accommodating scheduling practices that take attendees preferred meeting hours into account.

Through an analysis of the factors associated with time of day preferences, we found the need to commute to be significantly co-related with avoiding meetings at the edge of day. While, commuting to office was common for many information workers prior to the pandemic, we note that due to increased flexibility of work over time and space groupware tools that have an awareness of hybrid schedules (i.e. days that individuals work from the office versus from home) can result in better meeting scheduling recommendations. Further, scheduling tools can maximize opportunities for in-person collaborations by incorporating such awareness of individuals office schedules, a theme appreciated by some of our participants who strived to cluster their meetings on certain days to accommodate in-person meetings. 
Finally, we found one's job role and meeting load are strongly associated with their preferences for times of day for synchronous collaborations. 

\textbf{Meeting load:} As we saw throughout this work, meeting load is the single proximal driving factor in most scheduling practices, shaping their relationship with TOD, daily and weekly dispersion preferences. Users with lower meeting loads exhibit highly peaked distributions for times near the noon hour that the survey established as optimal times for many users. As meeting load increases (e.g. beyond 5 hours daily) the distribution of meetings shifts to be more even throughout the day. This pattern is likely driven by time constraints. Once a user has filled preferred time slots they must also fill less preferred times. 
Shifting our focus to Daily Dispersion patterns, we found that individuals broadly arrange their meetings in two different ways in an attempt to maintain productivity: clustering meetings in close temporal proximity (e.g. middle of the day) or spread them out to have buffer time before and after meetings.
Our mixed analysis of dispersion preferences and practices suggest that as a user’s meeting load rises, a user has choices for
meeting dispersion and thus we see variance in how spread-out meetings are across users. However, beyond a meeting load associated with about half of the workday, users choices and variation appear to decline. 
Given that, qualitatively, we see differences in the perceived desirability of arranging meeting and non-meeting work across the day in different information workers, our results highlights the negative impact of exceeding a certain threshold for daily meeting hours which can negatively impact individuals productivity and well-being.

Essentially, increased meeting load beyond an acceptable level (exceeding half of a workday) impacts individuals ability to distribute meeting load in a way that is aligned with their preferred rhythm of work, which can negatively impact their productivity and well-being.

Finally, our analysis of weekly dispersion patterns also indicated a strong relationship with meeting load: weekly dispersion, as measured by entropy, varies substantially with meeting load similarly to daily dispersion. Daily dispersion rises steeply with meeting load before quickly tapering off. This may reflect the challenges of packing an increasing number of meetings into a subset of days of the week. 

There are two main implications of these findings both for information workers productivity and for designing preference-aware scheduling systems. First, while past research \cite{yang2022effects} highlighted increased meeting load as a barrier to maintaining productivity, we empirically identified a threshold for meeting load that inhibits individuals ability to distribute meeting load in ways aligned with their productive rhythms.
Second, many intelligent systems (elsewhere and in scheduling domain) learn individual preferences from their behavioral signals (i.e. practices). Our results finds weak alignment between scheduling preferences and practices (w/ some notable exceptions, e.g., meeting organizers) and shows, empirically, why learning from practices of busy users (i.e. high meeting load) may not provide reliable signals for capturing individual scheduling preferences. This finding further motivates the need for methods that directly elicit scheduling and calendaring preferences and constrains from the users. 
With the recent advancements in conversational, LLM-powered interfaces integrated into our existing tools, we now have an excellent opportunity to gather individuals' scheduling preferences more directly and seamlessly within the context of their scheduling workflows.

\subsection{Design Implications for Calendar Scheduling Systems}
Our study provides several design implications for calendars scheduling system design. 
To make the scheduling system personalized, the system must have a holistic understanding of users' preferences. 
Similar to the PTime system~\cite{berry2007balancing}, systems should elicit  users' preferences along the four classes we identify. 

However, even after these preferences are identified, our findings on the limited alignment between preferences and practices indicate that a system that optimizes for meeting organizer's preferences may not be sufficient.
Many respondents suggest that increased visibility of attendee preferences to the organizer of a meeting may help improve overall scheduling outcomes:
\begin{quote}
    ``\textit{Seeing people's preferred times to meet would be very helpful.}'' [P72] \\
    ``\textit{It would help a lot to know what times that other participants are flexible. }'' [P96]
    ``\textit{}''
\end{quote}
As one participant notes, even with these suggestions, which require tradeoffs between privacy and transparency, it may be difficult to determine ``fair'' times: ``\textit{Finding a time that would make it "fair" is a pain. I spend a lot of time being considerate. I want a more objective way to decide how to potentially break someone's desired working/meeting hours.
}.'' P108.

By uncovering and understanding the dynamics of scheduling preferences, we can pave the way for developing systems that promote better individual and organizational outcomes in terms of productivity, well-being, and work-life balance. The identification of primary scheduling preferences serves as a foundational step towards designing systems that align with the needs and preferences of both organizers and attendees.
Future research can delve deeper into the balancing act of design criteria, considering factors such as the balance between transparency and privacy in scheduling systems, the trade-offs between organizational and individual outcomes, and the impact of different design choices on meeting effectiveness and efficiency. By addressing these considerations, we can strive to create scheduling systems that optimize the individual and organizational outcomes

%% file: 7-Limitation.tex
\section{Limitations}
Overall, we have endeavored to improve this study’s validity by combining worker behaviors, obtained from a large-scale trace data analysis, and worker perceptions, obtained from a survey with targeted questions about scheduling preferences. However, our results should have several limitations:
Firstly, the telemetry data we used to analyze practices and factors should be understood as behaviors around how people schedule meetings in the status quo, rather than explicit signals of people desiring certain calendar arrangements. Because this data is ambiguous about workers’ perceptions of their meetings, we sought to use our survey as a way of validating the patterns we observed in the data. 

Secondly, it is important to note that our study was conducted within a single corporation headquartered on the West Coast of the USA. This organizational context may introduce specific cultural and operational factors that influence meeting practices and preferences. Therefore, the generalizability of our findings to other organizations and geographical regions should be approached with caution. Businesses around the world are likely to exhibit significant heterogeneity in their scheduling practices and worker behaviors.

Lastly, our study acknowledges the limitation of the scale of survey responses. While we aimed to obtain a representative sample, a larger scale of survey responses could provide more comprehensive insights and potentially yield additional valuable findings regarding the reasoning behind preference categories and other factors influencing meeting practice.

%% file: Appendix.tex
\section{Appendix}
\section{Survey Consent Form}\label{appendix:consent}
\begin{quote}
\tiny
    Thank you for taking the time to consider volunteering in this experiment on meeting scheduling preferences and practices being conducted at COMPANY.  This form explains what will happen if you join this research project. Please read it carefully and take as much time as you need. Ask the study team about anything that is not clear.

    Participation in this study is voluntary and you will not be penalized if you decide not to take part in the study or if you quit the study later. 
    Before you start the study, here is some relevant information:\\

\textbf{Principal Investigators:}  ANONYMIZED\\
\textbf{PURPOSE:} The purpose of this project is to learn how organizers take on meeting scheduling workflows, what steps they follow and think about the challenges they face when scheduling events. Your responses will help us better understand how individuals’ scheduling practices are aligned with their time management preferences and how scheduling tools can facilitate different scheduling activities.\\
\textbf{PROCEDURES}\\
Throughout the survey, we will ask you to look at your work calendar in week view and respond to the questions. We expect this survey to take about 15 minutes to complete. Your responses will help us bring more intelligent scheduling experiences to Outlook. 

\textbf{PERSONAL INFORMATION AND CONFIDENTIALITY}\\
COMPANY is ultimately responsible for determining the purposes and uses of your personal information.\\ 

\textbf{Personal information we collect.} This survey collects your name and alias. We may use the alias to reach out if we need to do a follow-up after the survey.

\textbf{How we use personal information.}  The personal information and other data collected during this project will be used primarily to perform research for purposes described in the introduction above.  Such information and data, or the results of the research may eventually be used to develop and improve our internal policies and commercial products, services or technologies.

\textbf{How we store and share your personal information.}  The survey responses will be stored in a secured, limited-access location. Some people may need to look at your study information. They include: the researchers involved in this study, who may be COMPANY full time employees, fixed term employees, such as research interns, and the Privacy Team. The Privacy Team is a group that reviews the study to protect your rights as a research participant.   

\textbf{How you can access and control your personal information.}  If you wish to review or copy any personal information you provided during the study, or if you want us to delete or correct any such data, email your request to either of the Principal Investigators listed above.

\end{quote}

%% file: main.bbl

\begin{thebibliography}{89}


\ifx \showCODEN    \undefined \def \showCODEN     #1{\unskip}     \fi
\ifx \showDOI      \undefined \def \showDOI       #1{#1}\fi
\ifx \showISBNx    \undefined \def \showISBNx     #1{\unskip}     \fi
\ifx \showISBNxiii \undefined \def \showISBNxiii  #1{\unskip}     \fi
\ifx \showISSN     \undefined \def \showISSN      #1{\unskip}     \fi
\ifx \showLCCN     \undefined \def \showLCCN      #1{\unskip}     \fi
\ifx \shownote     \undefined \def \shownote      #1{#1}          \fi
\ifx \showarticletitle \undefined \def \showarticletitle #1{#1}   \fi
\ifx \showURL      \undefined \def \showURL       {\relax}        \fi
\providecommand\bibfield[2]{#2}
\providecommand\bibinfo[2]{#2}
\providecommand\natexlab[1]{#1}
\providecommand\showeprint[2][]{arXiv:#2}

\bibitem[\protect\citeauthoryear{??}{MST}{2021}]%
        {MSTriplePeak}
 \bibinfo{year}{2021}\natexlab{}.
\newblock \bibinfo{title}{The Rise of the Triple Peak Day. Microsoft 2021}.
\newblock
\newblock
\urldef\tempurl%
\url{https://www.microsoft.com/en-us/worklab/triple-peak-day}
\showURL{%
\tempurl}
\newblock
\shownote{Accessed: 2022-09-13.}


\bibitem[\protect\citeauthoryear{??}{MSW}{2022}]%
        {MSWTI}
 \bibinfo{year}{2022}\natexlab{}.
\newblock \bibinfo{title}{Great Expectations: Making Hybrid Work Work. Microsoft WorkLab: Work Trend Index 2022}.
\newblock
\newblock
\urldef\tempurl%
\url{https://www.microsoft.com/en-us/worklab/work-trend-index/great-expectations-making-hybrid-work-work}
\showURL{%
\tempurl}
\newblock
\shownote{Accessed: 2022-09-13.}


\bibitem[\protect\citeauthoryear{Agerfalk, Fitzgerald, Holmstrom~Olsson, Lings, Lundell, and {\'O}~Conch{\'u}ir}{Agerfalk et~al\mbox{.}}{2005}]%
        {agerfalk2005framework}
\bibfield{author}{\bibinfo{person}{Par~J Agerfalk}, \bibinfo{person}{Brian Fitzgerald}, \bibinfo{person}{Helena Holmstrom~Olsson}, \bibinfo{person}{Brian Lings}, \bibinfo{person}{Bjorn Lundell}, {and} \bibinfo{person}{Eoin {\'O}~Conch{\'u}ir}.} \bibinfo{year}{2005}\natexlab{}.
\newblock \showarticletitle{A framework for considering opportunities and threats in distributed software development}.
\newblock  (\bibinfo{year}{2005}).
\newblock


\bibitem[\protect\citeauthoryear{Ahrendt, Cabrita, Clerici, Hurley, Leon{\v{c}}ikas, Mascherini, Riso, and S{\'a}ndor}{Ahrendt et~al\mbox{.}}{2020}]%
        {ahrendt2020living}
\bibfield{author}{\bibinfo{person}{Daphne Ahrendt}, \bibinfo{person}{Jorge Cabrita}, \bibinfo{person}{Eleonora Clerici}, \bibinfo{person}{John Hurley}, \bibinfo{person}{Tadas Leon{\v{c}}ikas}, \bibinfo{person}{Massimiliano Mascherini}, \bibinfo{person}{Sara Riso}, {and} \bibinfo{person}{Eszter S{\'a}ndor}.} \bibinfo{year}{2020}\natexlab{}.
\newblock \showarticletitle{Living, working and COVID-19}.
\newblock  (\bibinfo{year}{2020}).
\newblock


\bibitem[\protect\citeauthoryear{Allen, Merlo, Lawrence, Slutsky, and Gray}{Allen et~al\mbox{.}}{2021}]%
        {allen2021boundary}
\bibfield{author}{\bibinfo{person}{Tammy~D Allen}, \bibinfo{person}{Kelsey Merlo}, \bibinfo{person}{Roxanne~C Lawrence}, \bibinfo{person}{Jeremiah Slutsky}, {and} \bibinfo{person}{Cheryl~E Gray}.} \bibinfo{year}{2021}\natexlab{}.
\newblock \showarticletitle{Boundary management and work-nonwork balance while working from home}.
\newblock \bibinfo{journal}{\emph{Applied Psychology}} \bibinfo{volume}{70}, \bibinfo{number}{1} (\bibinfo{year}{2021}), \bibinfo{pages}{60--84}.
\newblock


\bibitem[\protect\citeauthoryear{Awada, Lucas, Becerik-Gerber, and Roll}{Awada et~al\mbox{.}}{2021}]%
        {awada2021working}
\bibfield{author}{\bibinfo{person}{Mohamad Awada}, \bibinfo{person}{Gale Lucas}, \bibinfo{person}{Burcin Becerik-Gerber}, {and} \bibinfo{person}{Shawn Roll}.} \bibinfo{year}{2021}\natexlab{}.
\newblock \showarticletitle{Working from home during the COVID-19 pandemic: Impact on office worker productivity and work experience}.
\newblock \bibinfo{journal}{\emph{Work}} \bibinfo{number}{Preprint} (\bibinfo{year}{2021}), \bibinfo{pages}{1--19}.
\newblock


\bibitem[\protect\citeauthoryear{Ball}{Ball}{2010}]%
        {ball2010workplace}
\bibfield{author}{\bibinfo{person}{Kirstie Ball}.} \bibinfo{year}{2010}\natexlab{}.
\newblock \showarticletitle{Workplace surveillance: An overview}.
\newblock \bibinfo{journal}{\emph{Labor History}} \bibinfo{volume}{51}, \bibinfo{number}{1} (\bibinfo{year}{2010}), \bibinfo{pages}{87--106}.
\newblock


\bibitem[\protect\citeauthoryear{Batarseh, Usher, and Daspit}{Batarseh et~al\mbox{.}}{2017}]%
        {batarseh2017collaboration}
\bibfield{author}{\bibinfo{person}{Fadi~S Batarseh}, \bibinfo{person}{John~M Usher}, {and} \bibinfo{person}{Joshua~J Daspit}.} \bibinfo{year}{2017}\natexlab{}.
\newblock \showarticletitle{Collaboration capability in virtual teams: examining the influence on diversity and innovation}.
\newblock \bibinfo{journal}{\emph{International Journal of Innovation Management}} \bibinfo{volume}{21}, \bibinfo{number}{04} (\bibinfo{year}{2017}), \bibinfo{pages}{1750034}.
\newblock


\bibitem[\protect\citeauthoryear{Begole, Tang, and Hill}{Begole et~al\mbox{.}}{2003}]%
        {begole2003rhythm}
\bibfield{author}{\bibinfo{person}{James"~Bo" Begole}, \bibinfo{person}{John~C Tang}, {and} \bibinfo{person}{Rosco Hill}.} \bibinfo{year}{2003}\natexlab{}.
\newblock \showarticletitle{Rhythm modeling, visualizations and applications}. In \bibinfo{booktitle}{\emph{Proceedings of the 16th annual ACM symposium on User interface software and technology}}. \bibinfo{pages}{11--20}.
\newblock


\bibitem[\protect\citeauthoryear{Begole, Tang, Smith, and Yankelovich}{Begole et~al\mbox{.}}{2002}]%
        {begole2002work}
\bibfield{author}{\bibinfo{person}{James"~Bo" Begole}, \bibinfo{person}{John~C Tang}, \bibinfo{person}{Randall~B Smith}, {and} \bibinfo{person}{Nicole Yankelovich}.} \bibinfo{year}{2002}\natexlab{}.
\newblock \showarticletitle{Work rhythms: analyzing visualizations of awareness histories of distributed groups}. In \bibinfo{booktitle}{\emph{Proceedings of the 2002 ACM conference on Computer supported cooperative work}}. \bibinfo{pages}{334--343}.
\newblock


\bibitem[\protect\citeauthoryear{Berry, Gervasio, Peintner, and Yorke-Smith}{Berry et~al\mbox{.}}{2007}]%
        {berry2007balancing}
\bibfield{author}{\bibinfo{person}{Pauline Berry}, \bibinfo{person}{Melinda Gervasio}, \bibinfo{person}{Bart Peintner}, {and} \bibinfo{person}{Neil Yorke-Smith}.} \bibinfo{year}{2007}\natexlab{}.
\newblock \bibinfo{booktitle}{\emph{Balancing the needs of personalization and reasoning in a user-centric scheduling assistant}}.
\newblock \bibinfo{type}{{T}echnical {R}eport}. \bibinfo{institution}{SRI INTERNATIONAL MENLO PARK CA ARTIFICIAL INTELLIGENCE CENTER}.
\newblock


\bibitem[\protect\citeauthoryear{B{\o}dker and Gr{\"o}nvall}{B{\o}dker and Gr{\"o}nvall}{2013}]%
        {bodker2013calendars}
\bibfield{author}{\bibinfo{person}{Susanne B{\o}dker} {and} \bibinfo{person}{Erik Gr{\"o}nvall}.} \bibinfo{year}{2013}\natexlab{}.
\newblock \showarticletitle{Calendars: Time coordination and overview in families and beyond}. In \bibinfo{booktitle}{\emph{ECSCW 2013: Proceedings of the 13th European Conference on Computer Supported Cooperative Work, 21-25 September 2013, Paphos, Cyprus}}. Springer, \bibinfo{pages}{63--81}.
\newblock


\bibitem[\protect\citeauthoryear{Breideband, Talkad~Sukumar, Mark, Caruso, D'Mello, and Striegel}{Breideband et~al\mbox{.}}{2022}]%
        {breideband2022home}
\bibfield{author}{\bibinfo{person}{Thomas Breideband}, \bibinfo{person}{Poorna Talkad~Sukumar}, \bibinfo{person}{Gloria Mark}, \bibinfo{person}{Megan Caruso}, \bibinfo{person}{Sidney D'Mello}, {and} \bibinfo{person}{Aaron~D Striegel}.} \bibinfo{year}{2022}\natexlab{}.
\newblock \showarticletitle{Home-Life and Work Rhythm Diversity in Distributed Teamwork: A Study with Information Workers during the COVID-19 Pandemic}.
\newblock \bibinfo{journal}{\emph{Proceedings of the ACM on Human-Computer Interaction}} \bibinfo{volume}{6}, \bibinfo{number}{CSCW1} (\bibinfo{year}{2022}), \bibinfo{pages}{1--23}.
\newblock


\bibitem[\protect\citeauthoryear{Brzozowski, Carattini, Klemmer, Mihelich, Hu, and Ng}{Brzozowski et~al\mbox{.}}{2006}]%
        {brzozowski2006grouptime}
\bibfield{author}{\bibinfo{person}{Mike Brzozowski}, \bibinfo{person}{Kendra Carattini}, \bibinfo{person}{Scott~R Klemmer}, \bibinfo{person}{Patrick Mihelich}, \bibinfo{person}{Jiang Hu}, {and} \bibinfo{person}{Andrew~Y Ng}.} \bibinfo{year}{2006}\natexlab{}.
\newblock \showarticletitle{groupTime: preference based group scheduling}. In \bibinfo{booktitle}{\emph{Proceedings of the SIGCHI conference on Human Factors in computing systems}}. \bibinfo{pages}{1047--1056}.
\newblock


\bibitem[\protect\citeauthoryear{Cajochen, Kr{\"a}uchi, and Wirz-Justice}{Cajochen et~al\mbox{.}}{2003}]%
        {cajochen2003role}
\bibfield{author}{\bibinfo{person}{Christian Cajochen}, \bibinfo{person}{K Kr{\"a}uchi}, {and} \bibinfo{person}{A Wirz-Justice}.} \bibinfo{year}{2003}\natexlab{}.
\newblock \showarticletitle{Role of melatonin in the regulation of human circadian rhythms and sleep}.
\newblock \bibinfo{journal}{\emph{Journal of neuroendocrinology}} \bibinfo{volume}{15}, \bibinfo{number}{4} (\bibinfo{year}{2003}), \bibinfo{pages}{432--437}.
\newblock


\bibitem[\protect\citeauthoryear{Cambo, Avrahami, and Lee}{Cambo et~al\mbox{.}}{2017}]%
        {cambo2017breaksense}
\bibfield{author}{\bibinfo{person}{Scott~A Cambo}, \bibinfo{person}{Daniel Avrahami}, {and} \bibinfo{person}{Matthew~L Lee}.} \bibinfo{year}{2017}\natexlab{}.
\newblock \showarticletitle{BreakSense: Combining physiological and location sensing to promote mobility during work-breaks}. In \bibinfo{booktitle}{\emph{Proceedings of the 2017 CHI Conference on Human Factors in Computing Systems}}. \bibinfo{pages}{3595--3607}.
\newblock


\bibitem[\protect\citeauthoryear{Cao, Lee, Iqbal, Czerwinski, Wong, Rintel, Hecht, Teevan, and Yang}{Cao et~al\mbox{.}}{2021}]%
        {cao2021large}
\bibfield{author}{\bibinfo{person}{Hancheng Cao}, \bibinfo{person}{Chia-Jung Lee}, \bibinfo{person}{Shamsi Iqbal}, \bibinfo{person}{Mary Czerwinski}, \bibinfo{person}{Priscilla~NY Wong}, \bibinfo{person}{Sean Rintel}, \bibinfo{person}{Brent Hecht}, \bibinfo{person}{Jaime Teevan}, {and} \bibinfo{person}{Longqi Yang}.} \bibinfo{year}{2021}\natexlab{}.
\newblock \showarticletitle{Large scale analysis of multitasking behavior during remote meetings}. In \bibinfo{booktitle}{\emph{Proceedings of the 2021 CHI Conference on Human Factors in Computing Systems}}. \bibinfo{pages}{1--13}.
\newblock


\bibitem[\protect\citeauthoryear{Chajewska, Koller, and Parr}{Chajewska et~al\mbox{.}}{2000}]%
        {chajewska2000making}
\bibfield{author}{\bibinfo{person}{Urszula Chajewska}, \bibinfo{person}{Daphne Koller}, {and} \bibinfo{person}{Ronald Parr}.} \bibinfo{year}{2000}\natexlab{}.
\newblock \showarticletitle{Making rational decisions using adaptive utility elicitation}. In \bibinfo{booktitle}{\emph{Aaai/Iaai}}. \bibinfo{pages}{363--369}.
\newblock


\bibitem[\protect\citeauthoryear{Choi and Cho}{Choi and Cho}{2019}]%
        {choi2019mechanism}
\bibfield{author}{\bibinfo{person}{Ok-Kyu Choi} {and} \bibinfo{person}{Erin Cho}.} \bibinfo{year}{2019}\natexlab{}.
\newblock \showarticletitle{The mechanism of trust affecting collaboration in virtual teams and the moderating roles of the culture of autonomy and task complexity}.
\newblock \bibinfo{journal}{\emph{Computers in Human Behavior}}  \bibinfo{volume}{91} (\bibinfo{year}{2019}), \bibinfo{pages}{305--315}.
\newblock


\bibitem[\protect\citeauthoryear{Corbin and Strauss}{Corbin and Strauss}{1990}]%
        {corbin1990grounded}
\bibfield{author}{\bibinfo{person}{Juliet~M Corbin} {and} \bibinfo{person}{Anselm Strauss}.} \bibinfo{year}{1990}\natexlab{}.
\newblock \showarticletitle{Grounded theory research: Procedures, canons, and evaluative criteria}.
\newblock \bibinfo{journal}{\emph{Qualitative sociology}} \bibinfo{volume}{13}, \bibinfo{number}{1} (\bibinfo{year}{1990}), \bibinfo{pages}{3--21}.
\newblock


\bibitem[\protect\citeauthoryear{Cramton}{Cramton}{2001}]%
        {cramton2001mutual}
\bibfield{author}{\bibinfo{person}{Catherine~Durnell Cramton}.} \bibinfo{year}{2001}\natexlab{}.
\newblock \showarticletitle{The mutual knowledge problem and its consequences for dispersed collaboration}.
\newblock \bibinfo{journal}{\emph{Organization science}} \bibinfo{volume}{12}, \bibinfo{number}{3} (\bibinfo{year}{2001}), \bibinfo{pages}{346--371}.
\newblock


\bibitem[\protect\citeauthoryear{Cummings}{Cummings}{2011}]%
        {cummings2011geography}
\bibfield{author}{\bibinfo{person}{Jonathon~N Cummings}.} \bibinfo{year}{2011}\natexlab{}.
\newblock \showarticletitle{Geography is alive and well in virtual teams}.
\newblock \bibinfo{journal}{\emph{Commun. ACM}} \bibinfo{volume}{54}, \bibinfo{number}{8} (\bibinfo{year}{2011}), \bibinfo{pages}{24--26}.
\newblock


\bibitem[\protect\citeauthoryear{Cummings, Espinosa, and Pickering}{Cummings et~al\mbox{.}}{2009}]%
        {cummings2009crossing}
\bibfield{author}{\bibinfo{person}{Jonathon~N Cummings}, \bibinfo{person}{J~Alberto Espinosa}, {and} \bibinfo{person}{Cynthia~K Pickering}.} \bibinfo{year}{2009}\natexlab{}.
\newblock \showarticletitle{Crossing spatial and temporal boundaries in globally distributed projects: A relational model of coordination delay}.
\newblock \bibinfo{journal}{\emph{Information Systems Research}} \bibinfo{volume}{20}, \bibinfo{number}{3} (\bibinfo{year}{2009}), \bibinfo{pages}{420--439}.
\newblock


\bibitem[\protect\citeauthoryear{DeFilippis, Impink, Singell, Polzer, and Sadun}{DeFilippis et~al\mbox{.}}{2020}]%
        {defilippis2020collaborating}
\bibfield{author}{\bibinfo{person}{Evan DeFilippis}, \bibinfo{person}{Stephen~Michael Impink}, \bibinfo{person}{Madison Singell}, \bibinfo{person}{Jeffrey~T Polzer}, {and} \bibinfo{person}{Raffaella Sadun}.} \bibinfo{year}{2020}\natexlab{}.
\newblock \bibinfo{booktitle}{\emph{Collaborating during coronavirus: The impact of COVID-19 on the nature of work}}.
\newblock \bibinfo{type}{{T}echnical {R}eport}. \bibinfo{institution}{National Bureau of Economic Research}.
\newblock


\bibitem[\protect\citeauthoryear{Dennis, Fuller, and Valacich}{Dennis et~al\mbox{.}}{2008}]%
        {dennis2008media}
\bibfield{author}{\bibinfo{person}{Alan~R Dennis}, \bibinfo{person}{Robert~M Fuller}, {and} \bibinfo{person}{Joseph~S Valacich}.} \bibinfo{year}{2008}\natexlab{}.
\newblock \showarticletitle{Media, tasks, and communication processes: A theory of media synchronicity}.
\newblock \bibinfo{journal}{\emph{MIS quarterly}} (\bibinfo{year}{2008}), \bibinfo{pages}{575--600}.
\newblock


\bibitem[\protect\citeauthoryear{Dennis and Valacich}{Dennis and Valacich}{1999}]%
        {dennis1999rethinking}
\bibfield{author}{\bibinfo{person}{Alan~R Dennis} {and} \bibinfo{person}{Joseph~S Valacich}.} \bibinfo{year}{1999}\natexlab{}.
\newblock \showarticletitle{Rethinking media richness: Towards a theory of media synchronicity}. In \bibinfo{booktitle}{\emph{Proceedings of the 32nd Annual Hawaii International Conference on Systems Sciences. 1999. HICSS-32. Abstracts and CD-ROM of Full Papers}}. IEEE, \bibinfo{pages}{10--pp}.
\newblock


\bibitem[\protect\citeauthoryear{Dent, Boticario, McDermott, Mitchell, and Zabowski}{Dent et~al\mbox{.}}{1992}]%
        {dent1992personal}
\bibfield{author}{\bibinfo{person}{Lisa Dent}, \bibinfo{person}{Jesus Boticario}, \bibinfo{person}{John~P McDermott}, \bibinfo{person}{Tom~M Mitchell}, {and} \bibinfo{person}{David Zabowski}.} \bibinfo{year}{1992}\natexlab{}.
\newblock \showarticletitle{A personal learning apprentice}. In \bibinfo{booktitle}{\emph{AAAI}}. \bibinfo{pages}{96--103}.
\newblock


\bibitem[\protect\citeauthoryear{DeVault}{DeVault}{1991}]%
        {devault1991feeding}
\bibfield{author}{\bibinfo{person}{Marjorie~L DeVault}.} \bibinfo{year}{1991}\natexlab{}.
\newblock \bibinfo{booktitle}{\emph{Feeding the family: The social organization of caring as gendered work}}.
\newblock \bibinfo{publisher}{University of Chicago Press}.
\newblock


\bibitem[\protect\citeauthoryear{Dourish and Bellotti}{Dourish and Bellotti}{1992}]%
        {dourish1992awareness}
\bibfield{author}{\bibinfo{person}{Paul Dourish} {and} \bibinfo{person}{Victoria Bellotti}.} \bibinfo{year}{1992}\natexlab{}.
\newblock \showarticletitle{Awareness and coordination in shared workspaces}. In \bibinfo{booktitle}{\emph{Proceedings of the 1992 ACM conference on Computer-supported cooperative work}}. \bibinfo{pages}{107--114}.
\newblock


\bibitem[\protect\citeauthoryear{Dub{\'e} and Robey}{Dub{\'e} and Robey}{2009}]%
        {dube2009surviving}
\bibfield{author}{\bibinfo{person}{Line Dub{\'e}} {and} \bibinfo{person}{Daniel Robey}.} \bibinfo{year}{2009}\natexlab{}.
\newblock \showarticletitle{Surviving the paradoxes of virtual teamwork}.
\newblock \bibinfo{journal}{\emph{Information systems journal}} \bibinfo{volume}{19}, \bibinfo{number}{1} (\bibinfo{year}{2009}), \bibinfo{pages}{3--30}.
\newblock


\bibitem[\protect\citeauthoryear{Elliot and Carpendale}{Elliot and Carpendale}{2005}]%
        {elliot2005awareness}
\bibfield{author}{\bibinfo{person}{Kathryn Elliot} {and} \bibinfo{person}{Sheelagh Carpendale}.} \bibinfo{year}{2005}\natexlab{}.
\newblock \showarticletitle{Awareness and coordination: A calendar for families}.
\newblock  (\bibinfo{year}{2005}).
\newblock


\bibitem[\protect\citeauthoryear{Epstein, Avrahami, and Biehl}{Epstein et~al\mbox{.}}{2016}]%
        {epstein2016taking}
\bibfield{author}{\bibinfo{person}{Daniel~A Epstein}, \bibinfo{person}{Daniel Avrahami}, {and} \bibinfo{person}{Jacob~T Biehl}.} \bibinfo{year}{2016}\natexlab{}.
\newblock \showarticletitle{Taking 5: Work-breaks, productivity, and opportunities for personal informatics for knowledge workers}. In \bibinfo{booktitle}{\emph{Proceedings of the 2016 CHI Conference on Human Factors in Computing Systems}}. \bibinfo{pages}{673--684}.
\newblock


\bibitem[\protect\citeauthoryear{Espinosa, Cummings, and Pickering}{Espinosa et~al\mbox{.}}{2011}]%
        {espinosa2011time}
\bibfield{author}{\bibinfo{person}{J~Alberto Espinosa}, \bibinfo{person}{Jonathon~N Cummings}, {and} \bibinfo{person}{Cynthia Pickering}.} \bibinfo{year}{2011}\natexlab{}.
\newblock \showarticletitle{Time separation, coordination, and performance in technical teams}.
\newblock \bibinfo{journal}{\emph{IEEE Transactions on Engineering Management}} \bibinfo{volume}{59}, \bibinfo{number}{1} (\bibinfo{year}{2011}), \bibinfo{pages}{91--103}.
\newblock


\bibitem[\protect\citeauthoryear{Espinosa and Pickering}{Espinosa and Pickering}{2006}]%
        {espinosa2006effect}
\bibfield{author}{\bibinfo{person}{J~Alberto Espinosa} {and} \bibinfo{person}{Cynthia Pickering}.} \bibinfo{year}{2006}\natexlab{}.
\newblock \showarticletitle{The effect of time separation on coordination processes and outcomes: A case study}. In \bibinfo{booktitle}{\emph{Proceedings of the 39th Annual Hawaii International Conference on System Sciences (HICSS'06)}}, Vol.~\bibinfo{volume}{1}. IEEE, \bibinfo{pages}{25b--25b}.
\newblock


\bibitem[\protect\citeauthoryear{Folkard and Tucker}{Folkard and Tucker}{2003}]%
        {folkard2003shift}
\bibfield{author}{\bibinfo{person}{Simon Folkard} {and} \bibinfo{person}{Philip Tucker}.} \bibinfo{year}{2003}\natexlab{}.
\newblock \showarticletitle{Shift work, safety and productivity}.
\newblock \bibinfo{journal}{\emph{Occupational medicine}} \bibinfo{volume}{53}, \bibinfo{number}{2} (\bibinfo{year}{2003}), \bibinfo{pages}{95--101}.
\newblock


\bibitem[\protect\citeauthoryear{Gervasio, Moffitt, Pollack, Taylor, and Uribe}{Gervasio et~al\mbox{.}}{2005}]%
        {gervasio2005active}
\bibfield{author}{\bibinfo{person}{Melinda~T Gervasio}, \bibinfo{person}{Michael~D Moffitt}, \bibinfo{person}{Martha~E Pollack}, \bibinfo{person}{Joseph~M Taylor}, {and} \bibinfo{person}{Tomas~E Uribe}.} \bibinfo{year}{2005}\natexlab{}.
\newblock \showarticletitle{Active preference learning for personalized calendar scheduling assistance}. In \bibinfo{booktitle}{\emph{Proceedings of the 10th international conference on Intelligent user interfaces}}. \bibinfo{pages}{90--97}.
\newblock


\bibitem[\protect\citeauthoryear{Gibbs, Mengel, and Siemroth}{Gibbs et~al\mbox{.}}{2021}]%
        {gibbs2021work}
\bibfield{author}{\bibinfo{person}{Michael Gibbs}, \bibinfo{person}{Friederike Mengel}, {and} \bibinfo{person}{Christoph Siemroth}.} \bibinfo{year}{2021}\natexlab{}.
\newblock \showarticletitle{Work from home \& productivity: Evidence from personnel \& analytics data on IT professionals}.
\newblock \bibinfo{journal}{\emph{University of Chicago, Becker Friedman Institute for Economics Working Paper}} \bibinfo{number}{2021-56} (\bibinfo{year}{2021}).
\newblock


\bibitem[\protect\citeauthoryear{Gonz{\'a}lez and Mark}{Gonz{\'a}lez and Mark}{2004}]%
        {gonzalez2004constant}
\bibfield{author}{\bibinfo{person}{Victor~M Gonz{\'a}lez} {and} \bibinfo{person}{Gloria Mark}.} \bibinfo{year}{2004}\natexlab{}.
\newblock \showarticletitle{" Constant, constant, multi-tasking craziness" managing multiple working spheres}. In \bibinfo{booktitle}{\emph{Proceedings of the SIGCHI conference on Human factors in computing systems}}. \bibinfo{pages}{113--120}.
\newblock


\bibitem[\protect\citeauthoryear{Grimes and Brush}{Grimes and Brush}{2008}]%
        {grimes2008life}
\bibfield{author}{\bibinfo{person}{Andrea Grimes} {and} \bibinfo{person}{AJ Brush}.} \bibinfo{year}{2008}\natexlab{}.
\newblock \showarticletitle{Life scheduling to support multiple social roles}. In \bibinfo{booktitle}{\emph{Proceedings of the SIGCHI Conference on Human Factors in Computing Systems}}. \bibinfo{pages}{821--824}.
\newblock


\bibitem[\protect\citeauthoryear{H{\"a}kkil{\"a}, Karhu, Kalving, and Colley}{H{\"a}kkil{\"a} et~al\mbox{.}}{2020}]%
        {hakkila2020practical}
\bibfield{author}{\bibinfo{person}{Jonna H{\"a}kkil{\"a}}, \bibinfo{person}{Mari Karhu}, \bibinfo{person}{Matilda Kalving}, {and} \bibinfo{person}{Ashley Colley}.} \bibinfo{year}{2020}\natexlab{}.
\newblock \showarticletitle{Practical family challenges of remote schooling during COVID-19 pandemic in Finland}. In \bibinfo{booktitle}{\emph{Proceedings of the 11th Nordic conference on human-computer interaction: Shaping experiences, shaping society}}. \bibinfo{pages}{1--9}.
\newblock


\bibitem[\protect\citeauthoryear{Haynes, Sen, Arora, and Nadella}{Haynes et~al\mbox{.}}{1997}]%
        {haynes1997automated}
\bibfield{author}{\bibinfo{person}{Thomas Haynes}, \bibinfo{person}{Sandip Sen}, \bibinfo{person}{Neeraj Arora}, {and} \bibinfo{person}{Rajani Nadella}.} \bibinfo{year}{1997}\natexlab{}.
\newblock \showarticletitle{An automated meeting scheduling system that utilizes user preferences}. In \bibinfo{booktitle}{\emph{Proceedings of the First International Conference on Autonomous Agents}}. \bibinfo{pages}{308--315}.
\newblock


\bibitem[\protect\citeauthoryear{Healy, Dunstan, Salmon, Cerin, Shaw, Zimmet, and Owen}{Healy et~al\mbox{.}}{2008}]%
        {healy2008breaks}
\bibfield{author}{\bibinfo{person}{Genevieve~N Healy}, \bibinfo{person}{David~W Dunstan}, \bibinfo{person}{Jo Salmon}, \bibinfo{person}{Ester Cerin}, \bibinfo{person}{Jonathan~E Shaw}, \bibinfo{person}{Paul~Z Zimmet}, {and} \bibinfo{person}{Neville Owen}.} \bibinfo{year}{2008}\natexlab{}.
\newblock \showarticletitle{Breaks in sedentary time: beneficial associations with metabolic risk}.
\newblock \bibinfo{journal}{\emph{Diabetes care}} \bibinfo{volume}{31}, \bibinfo{number}{4} (\bibinfo{year}{2008}), \bibinfo{pages}{661--666}.
\newblock


\bibitem[\protect\citeauthoryear{Hongladarom}{Hongladarom}{2002}]%
        {hongladarom2002web}
\bibfield{author}{\bibinfo{person}{Soraj Hongladarom}.} \bibinfo{year}{2002}\natexlab{}.
\newblock \showarticletitle{The web of time and the dilemma of globalization}.
\newblock \bibinfo{journal}{\emph{The Information Society}} \bibinfo{volume}{18}, \bibinfo{number}{4} (\bibinfo{year}{2002}), \bibinfo{pages}{241--249}.
\newblock


\bibitem[\protect\citeauthoryear{Jackson, Ribes, Buyuktur, and Bowker}{Jackson et~al\mbox{.}}{2011}]%
        {jackson2011collaborative}
\bibfield{author}{\bibinfo{person}{Steven~J Jackson}, \bibinfo{person}{David Ribes}, \bibinfo{person}{Ayse Buyuktur}, {and} \bibinfo{person}{Geoffrey~C Bowker}.} \bibinfo{year}{2011}\natexlab{}.
\newblock \showarticletitle{Collaborative rhythm: temporal dissonance and alignment in collaborative scientific work}. In \bibinfo{booktitle}{\emph{Proceedings of the ACM 2011 conference on Computer supported cooperative work}}. \bibinfo{pages}{245--254}.
\newblock


\bibitem[\protect\citeauthoryear{Jalote, Palit, Kurien, and Peethamber}{Jalote et~al\mbox{.}}{2004}]%
        {jalote2004timeboxing}
\bibfield{author}{\bibinfo{person}{Pankaj Jalote}, \bibinfo{person}{Aveejeet Palit}, \bibinfo{person}{Priya Kurien}, {and} \bibinfo{person}{VT Peethamber}.} \bibinfo{year}{2004}\natexlab{}.
\newblock \showarticletitle{Timeboxing: a process model for iterative software development}.
\newblock \bibinfo{journal}{\emph{Journal of Systems and Software}} \bibinfo{volume}{70}, \bibinfo{number}{1-2} (\bibinfo{year}{2004}), \bibinfo{pages}{117--127}.
\newblock


\bibitem[\protect\citeauthoryear{Janb{\"o}cke, Gawlitta, D{\"o}rrenb{\"a}cher, and Hassenzahl}{Janb{\"o}cke et~al\mbox{.}}{2020}]%
        {janbocke2020finding}
\bibfield{author}{\bibinfo{person}{Sarah Janb{\"o}cke}, \bibinfo{person}{Alina Gawlitta}, \bibinfo{person}{Judith D{\"o}rrenb{\"a}cher}, {and} \bibinfo{person}{Marc Hassenzahl}.} \bibinfo{year}{2020}\natexlab{}.
\newblock \showarticletitle{Finding the inner clock: A chronobiology-based calendar}. In \bibinfo{booktitle}{\emph{Extended Abstracts of the 2020 CHI Conference on Human Factors in Computing Systems}}. \bibinfo{pages}{1--7}.
\newblock


\bibitem[\protect\citeauthoryear{Jun, McDuff, and Czerwinski}{Jun et~al\mbox{.}}{2019}]%
        {jun2019circadian}
\bibfield{author}{\bibinfo{person}{Eunice Jun}, \bibinfo{person}{Daniel McDuff}, {and} \bibinfo{person}{Mary Czerwinski}.} \bibinfo{year}{2019}\natexlab{}.
\newblock \showarticletitle{Circadian rhythms and physiological synchrony: Evidence of the impact of diversity on small group creativity}.
\newblock \bibinfo{journal}{\emph{Proceedings of the ACM on Human-Computer Interaction}} \bibinfo{volume}{3}, \bibinfo{number}{CSCW} (\bibinfo{year}{2019}), \bibinfo{pages}{1--22}.
\newblock


\bibitem[\protect\citeauthoryear{Kaur, Williams, McDuff, Czerwinski, Teevan, and Iqbal}{Kaur et~al\mbox{.}}{2020}]%
        {kaur2020optimizing}
\bibfield{author}{\bibinfo{person}{Harmanpreet Kaur}, \bibinfo{person}{Alex~C Williams}, \bibinfo{person}{Daniel McDuff}, \bibinfo{person}{Mary Czerwinski}, \bibinfo{person}{Jaime Teevan}, {and} \bibinfo{person}{Shamsi~T Iqbal}.} \bibinfo{year}{2020}\natexlab{}.
\newblock \showarticletitle{Optimizing for happiness and productivity: Modeling opportune moments for transitions and breaks at work}. In \bibinfo{booktitle}{\emph{Proceedings of the 2020 CHI Conference on Human Factors in Computing Systems}}. \bibinfo{pages}{1--15}.
\newblock


\bibitem[\protect\citeauthoryear{Kim, Lee, Choi, Choi, and Kang}{Kim et~al\mbox{.}}{2018}]%
        {kim2018learning}
\bibfield{author}{\bibinfo{person}{Donghyeon Kim}, \bibinfo{person}{Jinhyuk Lee}, \bibinfo{person}{Donghee Choi}, \bibinfo{person}{Jaehoon Choi}, {and} \bibinfo{person}{Jaewoo Kang}.} \bibinfo{year}{2018}\natexlab{}.
\newblock \showarticletitle{Learning user preferences and understanding calendar contexts for event scheduling}. In \bibinfo{booktitle}{\emph{Proceedings of the 27th ACM International Conference on Information and Knowledge Management}}. \bibinfo{pages}{337--346}.
\newblock


\bibitem[\protect\citeauthoryear{Kozierok and Maes}{Kozierok and Maes}{1993}]%
        {kozierok1993learning}
\bibfield{author}{\bibinfo{person}{Robyn Kozierok} {and} \bibinfo{person}{Pattie Maes}.} \bibinfo{year}{1993}\natexlab{}.
\newblock \showarticletitle{A learning interface agent for scheduling meetings}. In \bibinfo{booktitle}{\emph{Proceedings of the 1st international conference on Intelligent user interfaces}}. \bibinfo{pages}{81--88}.
\newblock


\bibitem[\protect\citeauthoryear{Krzywicki, Wobcke, and Wong}{Krzywicki et~al\mbox{.}}{2010}]%
        {krzywicki2010adaptive}
\bibfield{author}{\bibinfo{person}{Alfred Krzywicki}, \bibinfo{person}{Wayne Wobcke}, {and} \bibinfo{person}{Anna Wong}.} \bibinfo{year}{2010}\natexlab{}.
\newblock \showarticletitle{An adaptive calendar assistant using pattern mining for user preference modelling}. In \bibinfo{booktitle}{\emph{Proceedings of the 15th international conference on Intelligent user interfaces}}. \bibinfo{pages}{71--80}.
\newblock


\bibitem[\protect\citeauthoryear{Mark, Czerwinski, and Iqbal}{Mark et~al\mbox{.}}{2018}]%
        {mark2018effects}
\bibfield{author}{\bibinfo{person}{Gloria Mark}, \bibinfo{person}{Mary Czerwinski}, {and} \bibinfo{person}{Shamsi~T Iqbal}.} \bibinfo{year}{2018}\natexlab{}.
\newblock \showarticletitle{Effects of individual differences in blocking workplace distractions}. In \bibinfo{booktitle}{\emph{Proceedings of the 2018 CHI Conference on Human Factors in Computing Systems}}. \bibinfo{pages}{1--12}.
\newblock


\bibitem[\protect\citeauthoryear{Mark, Gudith, and Klocke}{Mark et~al\mbox{.}}{2008}]%
        {mark2008cost}
\bibfield{author}{\bibinfo{person}{Gloria Mark}, \bibinfo{person}{Daniela Gudith}, {and} \bibinfo{person}{Ulrich Klocke}.} \bibinfo{year}{2008}\natexlab{}.
\newblock \showarticletitle{The cost of interrupted work: more speed and stress}. In \bibinfo{booktitle}{\emph{Proceedings of the SIGCHI conference on Human Factors in Computing Systems}}. \bibinfo{pages}{107--110}.
\newblock


\bibitem[\protect\citeauthoryear{Mark, Iqbal, Czerwinski, and Johns}{Mark et~al\mbox{.}}{2014}]%
        {mark2014bored}
\bibfield{author}{\bibinfo{person}{Gloria Mark}, \bibinfo{person}{Shamsi~T Iqbal}, \bibinfo{person}{Mary Czerwinski}, {and} \bibinfo{person}{Paul Johns}.} \bibinfo{year}{2014}\natexlab{}.
\newblock \showarticletitle{Bored mondays and focused afternoons: the rhythm of attention and online activity in the workplace}. In \bibinfo{booktitle}{\emph{Proceedings of the SIGCHI Conference on Human Factors in Computing Systems}}. \bibinfo{pages}{3025--3034}.
\newblock


\bibitem[\protect\citeauthoryear{McTernan, Dollard, and LaMontagne}{McTernan et~al\mbox{.}}{2013}]%
        {mcternan2013depression}
\bibfield{author}{\bibinfo{person}{Wesley~P McTernan}, \bibinfo{person}{Maureen~F Dollard}, {and} \bibinfo{person}{Anthony~D LaMontagne}.} \bibinfo{year}{2013}\natexlab{}.
\newblock \showarticletitle{Depression in the workplace: An economic cost analysis of depression-related productivity loss attributable to job strain and bullying}.
\newblock \bibinfo{journal}{\emph{Work \& Stress}} \bibinfo{volume}{27}, \bibinfo{number}{4} (\bibinfo{year}{2013}), \bibinfo{pages}{321--338}.
\newblock


\bibitem[\protect\citeauthoryear{Mell, Jang, and Chai}{Mell et~al\mbox{.}}{2021}]%
        {mell2021bridging}
\bibfield{author}{\bibinfo{person}{Julija~N Mell}, \bibinfo{person}{Sujin Jang}, {and} \bibinfo{person}{Sen Chai}.} \bibinfo{year}{2021}\natexlab{}.
\newblock \showarticletitle{Bridging temporal divides: Temporal brokerage in global teams and its impact on individual performance}.
\newblock \bibinfo{journal}{\emph{Organization Science}} \bibinfo{volume}{32}, \bibinfo{number}{3} (\bibinfo{year}{2021}), \bibinfo{pages}{731--751}.
\newblock


\bibitem[\protect\citeauthoryear{Meyer, Barr, Bird, and Zimmermann}{Meyer et~al\mbox{.}}{2019}]%
        {meyer2019today}
\bibfield{author}{\bibinfo{person}{Andr{\'e}~N Meyer}, \bibinfo{person}{Earl~T Barr}, \bibinfo{person}{Christian Bird}, {and} \bibinfo{person}{Thomas Zimmermann}.} \bibinfo{year}{2019}\natexlab{}.
\newblock \showarticletitle{Today was a good day: The daily life of software developers}.
\newblock \bibinfo{journal}{\emph{IEEE Transactions on Software Engineering}} \bibinfo{volume}{47}, \bibinfo{number}{5} (\bibinfo{year}{2019}), \bibinfo{pages}{863--880}.
\newblock


\bibitem[\protect\citeauthoryear{Meyer, Barton, Murphy, Zimmermann, and Fritz}{Meyer et~al\mbox{.}}{2017}]%
        {meyer2017work}
\bibfield{author}{\bibinfo{person}{Andr{\'e}~N Meyer}, \bibinfo{person}{Laura~E Barton}, \bibinfo{person}{Gail~C Murphy}, \bibinfo{person}{Thomas Zimmermann}, {and} \bibinfo{person}{Thomas Fritz}.} \bibinfo{year}{2017}\natexlab{}.
\newblock \showarticletitle{The work life of developers: Activities, switches and perceived productivity}.
\newblock \bibinfo{journal}{\emph{IEEE Transactions on Software Engineering}} \bibinfo{volume}{43}, \bibinfo{number}{12} (\bibinfo{year}{2017}), \bibinfo{pages}{1178--1193}.
\newblock


\bibitem[\protect\citeauthoryear{Mihaly}{Mihaly}{1997}]%
        {mihaly1997finding}
\bibfield{author}{\bibinfo{person}{Csikszentmihalyi Mihaly}.} \bibinfo{year}{1997}\natexlab{}.
\newblock \bibinfo{title}{Finding flow: The psychology of engagement with everyday life}.
\newblock
\newblock


\bibitem[\protect\citeauthoryear{Mitchell, Caruana, Freitag, McDermott, Zabowski, et~al\mbox{.}}{Mitchell et~al\mbox{.}}{1994}]%
        {mitchell1994experience}
\bibfield{author}{\bibinfo{person}{Tom~M Mitchell}, \bibinfo{person}{Rich Caruana}, \bibinfo{person}{Dayne Freitag}, \bibinfo{person}{John McDermott}, \bibinfo{person}{David Zabowski}, {et~al\mbox{.}}} \bibinfo{year}{1994}\natexlab{}.
\newblock \showarticletitle{Experience with a learning personal assistant}.
\newblock \bibinfo{journal}{\emph{Commun. ACM}} \bibinfo{volume}{37}, \bibinfo{number}{7} (\bibinfo{year}{1994}), \bibinfo{pages}{80--91}.
\newblock


\bibitem[\protect\citeauthoryear{Mok, Sun, Sen, and Sarrafzadeh}{Mok et~al\mbox{.}}{2023}]%
        {mok2023challenging}
\bibfield{author}{\bibinfo{person}{Lillio Mok}, \bibinfo{person}{Lu Sun}, \bibinfo{person}{Shilad Sen}, {and} \bibinfo{person}{Bahareh Sarrafzadeh}.} \bibinfo{year}{2023}\natexlab{}.
\newblock \showarticletitle{Challenging but Connective: Large-Scale Characteristics of Synchronous Collaboration Across Time Zones}. In \bibinfo{booktitle}{\emph{Proceedings of the 2023 CHI Conference on Human Factors in Computing Systems}}. \bibinfo{pages}{1--17}.
\newblock


\bibitem[\protect\citeauthoryear{Morrison-Smith and Ruiz}{Morrison-Smith and Ruiz}{2020}]%
        {morrison2020challenges}
\bibfield{author}{\bibinfo{person}{Sarah Morrison-Smith} {and} \bibinfo{person}{Jaime Ruiz}.} \bibinfo{year}{2020}\natexlab{}.
\newblock \showarticletitle{Challenges and barriers in virtual teams: a literature review}.
\newblock \bibinfo{journal}{\emph{SN Applied Sciences}} \bibinfo{volume}{2}, \bibinfo{number}{6} (\bibinfo{year}{2020}), \bibinfo{pages}{1--33}.
\newblock


\bibitem[\protect\citeauthoryear{Morshed, Hernandez, McDuff, Suh, Howe, Rowan, Abdin, Ramos, Tran, and Czerwinski}{Morshed et~al\mbox{.}}{2022}]%
        {morshed2022stress}
\bibfield{author}{\bibinfo{person}{M. Morshed}, \bibinfo{person}{J. Hernandez}, \bibinfo{person}{D. McDuff}, \bibinfo{person}{J. Suh}, \bibinfo{person}{E. Howe}, \bibinfo{person}{K. Rowan}, \bibinfo{person}{M. Abdin}, \bibinfo{person}{G. Ramos}, \bibinfo{person}{T. Tran}, {and} \bibinfo{person}{M. Czerwinski}.} \bibinfo{year}{2022}\natexlab{}.
\newblock \showarticletitle{Advancing the Understanding and Measurement of Workplace Stress in Remote Information Workers from Passive Sensors and Behavioral Data}. In \bibinfo{booktitle}{\emph{Proceedings of 10th International Conference on Affective Computing and Intelligent Interaction (ACII)}}.
\newblock


\bibitem[\protect\citeauthoryear{Murray and Harvey}{Murray and Harvey}{2010}]%
        {murray2010circadian}
\bibfield{author}{\bibinfo{person}{Greg Murray} {and} \bibinfo{person}{Allison Harvey}.} \bibinfo{year}{2010}\natexlab{}.
\newblock \showarticletitle{Circadian rhythms and sleep in bipolar disorder}.
\newblock \bibinfo{journal}{\emph{Bipolar disorders}} \bibinfo{volume}{12}, \bibinfo{number}{5} (\bibinfo{year}{2010}), \bibinfo{pages}{459--472}.
\newblock


\bibitem[\protect\citeauthoryear{Neeley}{Neeley}{2015}]%
        {neeley2015global}
\bibfield{author}{\bibinfo{person}{Tsedal Neeley}.} \bibinfo{year}{2015}\natexlab{}.
\newblock \showarticletitle{Global teams that work}.
\newblock \bibinfo{journal}{\emph{Harvard Business Review}} \bibinfo{volume}{93}, \bibinfo{number}{10} (\bibinfo{year}{2015}), \bibinfo{pages}{74--81}.
\newblock


\bibitem[\protect\citeauthoryear{Neustaedter, Brush, and Greenberg}{Neustaedter et~al\mbox{.}}{2009}]%
        {neustaedter2009calendar}
\bibfield{author}{\bibinfo{person}{Carman Neustaedter}, \bibinfo{person}{AJ~Bernheim Brush}, {and} \bibinfo{person}{Saul Greenberg}.} \bibinfo{year}{2009}\natexlab{}.
\newblock \showarticletitle{The calendar is crucial: Coordination and awareness through the family calendar}.
\newblock \bibinfo{journal}{\emph{ACM Transactions on Computer-Human Interaction (TOCHI)}} \bibinfo{volume}{16}, \bibinfo{number}{1} (\bibinfo{year}{2009}), \bibinfo{pages}{1--48}.
\newblock


\bibitem[\protect\citeauthoryear{Nissenbaum}{Nissenbaum}{2004}]%
        {nissenbaum2004privacy}
\bibfield{author}{\bibinfo{person}{Helen Nissenbaum}.} \bibinfo{year}{2004}\natexlab{}.
\newblock \showarticletitle{Privacy as contextual integrity}.
\newblock \bibinfo{journal}{\emph{Wash. L. Rev.}}  \bibinfo{volume}{79} (\bibinfo{year}{2004}), \bibinfo{pages}{119}.
\newblock


\bibitem[\protect\citeauthoryear{Oh and Smith}{Oh and Smith}{2005}]%
        {oh2005calendar}
\bibfield{author}{\bibinfo{person}{Jean Oh} {and} \bibinfo{person}{Stephen~F Smith}.} \bibinfo{year}{2005}\natexlab{}.
\newblock \showarticletitle{Calendar Assistants that Learn Preferences.}. In \bibinfo{booktitle}{\emph{AAAI Spring Symposium: Persistent Assistants: Living and Working with AI}}. \bibinfo{pages}{7--13}.
\newblock


\bibitem[\protect\citeauthoryear{Olson and Olson}{Olson and Olson}{2000}]%
        {olson2000distance}
\bibfield{author}{\bibinfo{person}{Gary~M Olson} {and} \bibinfo{person}{Judith~S Olson}.} \bibinfo{year}{2000}\natexlab{}.
\newblock \showarticletitle{Distance matters}.
\newblock \bibinfo{journal}{\emph{Human--computer interaction}} \bibinfo{volume}{15}, \bibinfo{number}{2-3} (\bibinfo{year}{2000}), \bibinfo{pages}{139--178}.
\newblock


\bibitem[\protect\citeauthoryear{R{\"a}ih{\"a}, Nerg, Jurvelin, Conlin, Korhonen, and Ala-Mursula}{R{\"a}ih{\"a} et~al\mbox{.}}{2021}]%
        {raiha2021evening}
\bibfield{author}{\bibinfo{person}{Tapio R{\"a}ih{\"a}}, \bibinfo{person}{Iiro Nerg}, \bibinfo{person}{Heidi Jurvelin}, \bibinfo{person}{Andrew Conlin}, \bibinfo{person}{Marko Korhonen}, {and} \bibinfo{person}{Leena Ala-Mursula}.} \bibinfo{year}{2021}\natexlab{}.
\newblock \showarticletitle{Evening chronotype is associated with poor work ability and disability pensions at midlife: a Northern Finland Birth Cohort 1966 Study}.
\newblock \bibinfo{journal}{\emph{Occupational and Environmental Medicine}} \bibinfo{volume}{78}, \bibinfo{number}{8} (\bibinfo{year}{2021}), \bibinfo{pages}{567--575}.
\newblock


\bibitem[\protect\citeauthoryear{Reddy and Dourish}{Reddy and Dourish}{2002}]%
        {reddy2002finger}
\bibfield{author}{\bibinfo{person}{Madhu Reddy} {and} \bibinfo{person}{Paul Dourish}.} \bibinfo{year}{2002}\natexlab{}.
\newblock \showarticletitle{A finger on the pulse: temporal rhythms and information seeking in medical work}. In \bibinfo{booktitle}{\emph{Proceedings of the 2002 ACM conference on Computer supported cooperative work}}. \bibinfo{pages}{344--353}.
\newblock


\bibitem[\protect\citeauthoryear{Romero, Reinecke, and Robert~Jr}{Romero et~al\mbox{.}}{2017}]%
        {romero2017influence}
\bibfield{author}{\bibinfo{person}{Daniel~M Romero}, \bibinfo{person}{Katharina Reinecke}, {and} \bibinfo{person}{Lionel~P Robert~Jr}.} \bibinfo{year}{2017}\natexlab{}.
\newblock \showarticletitle{The influence of early respondents: information cascade effects in online event scheduling}. In \bibinfo{booktitle}{\emph{Proceedings of the Tenth ACM International Conference on Web Search and Data Mining}}. \bibinfo{pages}{101--110}.
\newblock


\bibitem[\protect\citeauthoryear{Sarker, Ahuja, Sarker, and Kirkeby}{Sarker et~al\mbox{.}}{2011}]%
        {sarker2011role}
\bibfield{author}{\bibinfo{person}{Saonee Sarker}, \bibinfo{person}{Manju Ahuja}, \bibinfo{person}{Suprateek Sarker}, {and} \bibinfo{person}{Sarah Kirkeby}.} \bibinfo{year}{2011}\natexlab{}.
\newblock \showarticletitle{The role of communication and trust in global virtual teams: A social network perspective}.
\newblock \bibinfo{journal}{\emph{Journal of Management Information Systems}} \bibinfo{volume}{28}, \bibinfo{number}{1} (\bibinfo{year}{2011}), \bibinfo{pages}{273--310}.
\newblock


\bibitem[\protect\citeauthoryear{Schmidt, Collette, Cajochen, and Peigneux}{Schmidt et~al\mbox{.}}{2007}]%
        {schmidt2007time}
\bibfield{author}{\bibinfo{person}{Christina Schmidt}, \bibinfo{person}{Fabienne Collette}, \bibinfo{person}{Christian Cajochen}, {and} \bibinfo{person}{Philippe Peigneux}.} \bibinfo{year}{2007}\natexlab{}.
\newblock \showarticletitle{A time to think: circadian rhythms in human cognition}.
\newblock \bibinfo{journal}{\emph{Cognitive neuropsychology}} \bibinfo{volume}{24}, \bibinfo{number}{7} (\bibinfo{year}{2007}), \bibinfo{pages}{755--789}.
\newblock


\bibitem[\protect\citeauthoryear{Sen, Haynes, and Arora}{Sen et~al\mbox{.}}{1997}]%
        {sen1997satisfying}
\bibfield{author}{\bibinfo{person}{Sandip Sen}, \bibinfo{person}{Thomas Haynes}, {and} \bibinfo{person}{Neeraj Arora}.} \bibinfo{year}{1997}\natexlab{}.
\newblock \showarticletitle{Satisfying user preferences while negotiating meetings}.
\newblock \bibinfo{journal}{\emph{International Journal of Human-Computer Studies}} \bibinfo{volume}{47}, \bibinfo{number}{3} (\bibinfo{year}{1997}), \bibinfo{pages}{407--427}.
\newblock


\bibitem[\protect\citeauthoryear{Shifrin and Michel}{Shifrin and Michel}{2022}]%
        {shifrin2022flexible}
\bibfield{author}{\bibinfo{person}{Nicole~V Shifrin} {and} \bibinfo{person}{Jesse~S Michel}.} \bibinfo{year}{2022}\natexlab{}.
\newblock \showarticletitle{Flexible work arrangements and employee health: A meta-analytic review}.
\newblock \bibinfo{journal}{\emph{Work \& Stress}} \bibinfo{volume}{36}, \bibinfo{number}{1} (\bibinfo{year}{2022}), \bibinfo{pages}{60--85}.
\newblock


\bibitem[\protect\citeauthoryear{Steinhardt and Jackson}{Steinhardt and Jackson}{2014}]%
        {steinhardt2014reconciling}
\bibfield{author}{\bibinfo{person}{Stephanie~B Steinhardt} {and} \bibinfo{person}{Steven~J Jackson}.} \bibinfo{year}{2014}\natexlab{}.
\newblock \showarticletitle{Reconciling rhythms: plans and temporal alignment in collaborative scientific work}. In \bibinfo{booktitle}{\emph{Proceedings of the 17th ACM conference on Computer supported cooperative work \& social computing}}. \bibinfo{pages}{134--145}.
\newblock


\bibitem[\protect\citeauthoryear{Stone, Hedges, Neale, and Satin}{Stone et~al\mbox{.}}{1985}]%
        {stone1985prospective}
\bibfield{author}{\bibinfo{person}{Arthur~A Stone}, \bibinfo{person}{Susan~M Hedges}, \bibinfo{person}{John~M Neale}, {and} \bibinfo{person}{Maurice~S Satin}.} \bibinfo{year}{1985}\natexlab{}.
\newblock \showarticletitle{Prospective and cross-sectional mood reports offer no evidence of a" blue Monday" phenomenon.}
\newblock \bibinfo{journal}{\emph{Journal of Personality and Social Psychology}} \bibinfo{volume}{49}, \bibinfo{number}{1} (\bibinfo{year}{1985}), \bibinfo{pages}{129}.
\newblock


\bibitem[\protect\citeauthoryear{Strongman and Burt}{Strongman and Burt}{2000}]%
        {strongman2000taking}
\bibfield{author}{\bibinfo{person}{Kenneth~T Strongman} {and} \bibinfo{person}{Christopher~DB Burt}.} \bibinfo{year}{2000}\natexlab{}.
\newblock \showarticletitle{Taking breaks from work: An exploratory inquiry}.
\newblock \bibinfo{journal}{\emph{The Journal of psychology}} \bibinfo{volume}{134}, \bibinfo{number}{3} (\bibinfo{year}{2000}), \bibinfo{pages}{229--242}.
\newblock


\bibitem[\protect\citeauthoryear{Syakur, Khotimah, Rochman, and Satoto}{Syakur et~al\mbox{.}}{2018}]%
        {syakur2018integration}
\bibfield{author}{\bibinfo{person}{MA Syakur}, \bibinfo{person}{BK Khotimah}, \bibinfo{person}{EMS Rochman}, {and} \bibinfo{person}{Budi~Dwi Satoto}.} \bibinfo{year}{2018}\natexlab{}.
\newblock \showarticletitle{Integration k-means clustering method and elbow method for identification of the best customer profile cluster}. In \bibinfo{booktitle}{\emph{IOP conference series: materials science and engineering}}, Vol.~\bibinfo{volume}{336}. IOP Publishing, \bibinfo{pages}{012017}.
\newblock


\bibitem[\protect\citeauthoryear{Tang, Yankelovich, Begole, Van~Kleek, Li, and Bhalodia}{Tang et~al\mbox{.}}{2001}]%
        {tang2001connexus}
\bibfield{author}{\bibinfo{person}{John~C Tang}, \bibinfo{person}{Nicole Yankelovich}, \bibinfo{person}{James Begole}, \bibinfo{person}{Max Van~Kleek}, \bibinfo{person}{Francis Li}, {and} \bibinfo{person}{Janak Bhalodia}.} \bibinfo{year}{2001}\natexlab{}.
\newblock \showarticletitle{ConNexus to Awarenex: Extending awareness to mobile users}. In \bibinfo{booktitle}{\emph{Proceedings of the SIGCHI conference on Human factors in computing systems}}. \bibinfo{pages}{221--228}.
\newblock


\bibitem[\protect\citeauthoryear{Tang, Zhao, Cao, and Inkpen}{Tang et~al\mbox{.}}{2011}]%
        {tang2011your}
\bibfield{author}{\bibinfo{person}{John~C Tang}, \bibinfo{person}{Chen Zhao}, \bibinfo{person}{Xiang Cao}, {and} \bibinfo{person}{Kori Inkpen}.} \bibinfo{year}{2011}\natexlab{}.
\newblock \showarticletitle{Your time zone or mine? A study of globally time zone-shifted collaboration}. In \bibinfo{booktitle}{\emph{Proceedings of the ACM 2011 conference on Computer supported cooperative work}}. \bibinfo{pages}{235--244}.
\newblock


\bibitem[\protect\citeauthoryear{Teevan, Baym, Butler, Hecht, Jaffe, Nowak, Sellen, and Yang}{Teevan et~al\mbox{.}}{2022}]%
        {teevan2022microsoft}
\bibfield{author}{\bibinfo{person}{J Teevan}, \bibinfo{person}{N Baym}, \bibinfo{person}{J Butler}, \bibinfo{person}{B Hecht}, \bibinfo{person}{S Jaffe}, \bibinfo{person}{K Nowak}, \bibinfo{person}{A Sellen}, {and} \bibinfo{person}{L Yang}.} \bibinfo{year}{2022}\natexlab{}.
\newblock \bibinfo{booktitle}{\emph{Microsoft new future of work report 2022}}.
\newblock \bibinfo{type}{{T}echnical {R}eport}. \bibinfo{institution}{Technical Report. Microsoft Research Tech Report MSR-TR-2022--3}.
\newblock


\bibitem[\protect\citeauthoryear{Tietze and Musson}{Tietze and Musson}{2003}]%
        {tietze2003times}
\bibfield{author}{\bibinfo{person}{Susanne Tietze} {and} \bibinfo{person}{Gill Musson}.} \bibinfo{year}{2003}\natexlab{}.
\newblock \showarticletitle{The times and temporalities of home-based telework}.
\newblock \bibinfo{journal}{\emph{Personnel Review}} \bibinfo{volume}{32}, \bibinfo{number}{4} (\bibinfo{year}{2003}), \bibinfo{pages}{438--455}.
\newblock


\bibitem[\protect\citeauthoryear{Tullio, Goecks, Mynatt, and Nguyen}{Tullio et~al\mbox{.}}{2002}]%
        {tullio2002augmenting}
\bibfield{author}{\bibinfo{person}{Joe Tullio}, \bibinfo{person}{Jeremy Goecks}, \bibinfo{person}{Elizabeth~D Mynatt}, {and} \bibinfo{person}{David~H Nguyen}.} \bibinfo{year}{2002}\natexlab{}.
\newblock \showarticletitle{Augmenting shared personal calendars}. In \bibinfo{booktitle}{\emph{Proceedings of the 15th annual ACM symposium on User interface software and technology}}. \bibinfo{pages}{11--20}.
\newblock


\bibitem[\protect\citeauthoryear{Viappiani, Faltings, and Pu}{Viappiani et~al\mbox{.}}{2006}]%
        {viappiani2006evaluating}
\bibfield{author}{\bibinfo{person}{Paolo Viappiani}, \bibinfo{person}{Boi Faltings}, {and} \bibinfo{person}{Pearl Pu}.} \bibinfo{year}{2006}\natexlab{}.
\newblock \showarticletitle{Evaluating preference-based search tools: a tale of two approaches}. In \bibinfo{booktitle}{\emph{Proceedings of the Twenty-first National Conference on Artificial Intelligence (AAAI-06)}}. AAAI press, \bibinfo{pages}{205--211}.
\newblock


\bibitem[\protect\citeauthoryear{Wieth and Zacks}{Wieth and Zacks}{2011}]%
        {wieth2011time}
\bibfield{author}{\bibinfo{person}{Mareike~B Wieth} {and} \bibinfo{person}{Rose~T Zacks}.} \bibinfo{year}{2011}\natexlab{}.
\newblock \showarticletitle{Time of day effects on problem solving: When the non-optimal is optimal}.
\newblock \bibinfo{journal}{\emph{Thinking \& Reasoning}} \bibinfo{volume}{17}, \bibinfo{number}{4} (\bibinfo{year}{2011}), \bibinfo{pages}{387--401}.
\newblock


\bibitem[\protect\citeauthoryear{Williams, Iqbal, Kiseleva, and White}{Williams et~al\mbox{.}}{2023}]%
        {williams2023managing}
\bibfield{author}{\bibinfo{person}{Alex~C Williams}, \bibinfo{person}{Shamsi Iqbal}, \bibinfo{person}{Julia Kiseleva}, {and} \bibinfo{person}{Ryen~W White}.} \bibinfo{year}{2023}\natexlab{}.
\newblock \showarticletitle{Managing Tasks across the Work--Life Boundary: Opportunities, Challenges, and Directions}.
\newblock \bibinfo{journal}{\emph{ACM Transactions on Computer-Human Interaction}} \bibinfo{volume}{30}, \bibinfo{number}{3} (\bibinfo{year}{2023}), \bibinfo{pages}{1--31}.
\newblock


\bibitem[\protect\citeauthoryear{Yang, Holtz, Jaffe, Suri, Sinha, Weston, Joyce, Shah, Sherman, Hecht, et~al\mbox{.}}{Yang et~al\mbox{.}}{2022}]%
        {yang2022effects}
\bibfield{author}{\bibinfo{person}{Longqi Yang}, \bibinfo{person}{David Holtz}, \bibinfo{person}{Sonia Jaffe}, \bibinfo{person}{Siddharth Suri}, \bibinfo{person}{Shilpi Sinha}, \bibinfo{person}{Jeffrey Weston}, \bibinfo{person}{Connor Joyce}, \bibinfo{person}{Neha Shah}, \bibinfo{person}{Kevin Sherman}, \bibinfo{person}{Brent Hecht}, {et~al\mbox{.}}} \bibinfo{year}{2022}\natexlab{}.
\newblock \showarticletitle{The effects of remote work on collaboration among information workers}.
\newblock \bibinfo{journal}{\emph{Nature human behaviour}} \bibinfo{volume}{6}, \bibinfo{number}{1} (\bibinfo{year}{2022}), \bibinfo{pages}{43--54}.
\newblock


\end{thebibliography}
